\documentclass[aps,prd,showpacs,amsmath,amssymb,nobibnotes,nofootinbib,11pt]{revtex4-1}
\usepackage{appendix}
\usepackage{hyperref}
\usepackage{graphicx}
\usepackage{algorithm}
\usepackage{algorithmic}
\usepackage{amsmath}
\usepackage{txfonts}
\usepackage{subfigure}
\usepackage{amssymb}
\usepackage{subfigure}
\usepackage{multirow}
\usepackage{youngtab}
\usepackage{colortbl}
\usepackage{slashbox}

\usepackage{hhline}
\newtheorem{theorem}{Theorem}[section]
\newtheorem{definition}[theorem]{Definition}

\newenvironment{defn*}{\begin{definition}}{\end{definition}}

\begin{document}
\title{Quantum Mechanics on Curved Hypersurfaces}

\author{Mehmet Ali Olpak}
\email{maliolpak@gmail.com}
\affiliation{Department of Physics, Middle East Technical University,
Ankara, Turkey}

\begin{abstract}
In this work, Schr\"odinger and Dirac equations will be examined in geometries that confine the particles to hypersurfaces. For this purpose, two methods will be considered. The first method is the thin layer method which relies on explicit use of geometrical relations and the squeezing of a certain coordinate of space (or spacetime). The second is Dirac's quantization procedure involving the modification of canonical quantization making use of the geometrical constraints. For the Dirac equation, only the first method will be considered. Lastly, the results of the two methods will be compared and some notes on the differences between the results will be included.  
\end{abstract}

\maketitle
\tableofcontents
\newpage

\section{INTRODUCTION}
\label{sec:introduction}

Today, there are many problems concerning systems having spacetime dimensionality less than four. These problems involve a wide variety of topics ranging from condensed matter physics to gravitation. Some systems treated within this context are, however, only effectively three or two dimensional while indeed living in four spacetime dimensions. As an example, one can think of electrons within a carbon nanotube having a single layer of carbon atoms \cite{Ferrari}. These electrons are under the influence of the underlying lattice, which is actually curved. At this point, one can ask the following questions: Does the geometry of the lattice affect the motion of the electrons? If it does, how?

Such questions can be answered by using a suitable method which takes care of the geometry of the system. We already know how to do this; the quantum mechanical central force problem is an example. If we move one step further, and change the picture a little bit, more interesting things happen. Consider the same example, the carbon nanotube, which has, say, a cylindrical shape, and let the length of the tube be much greater than its thickness. Then, one can treat the electrons as if they are \textquoteleft constrained\textquoteright \, to move on a cylindrical surface. So, how can this picture be viewed as the description of a constrained system, which obeys the laws of quantum mechanics? This will be the main question that is to be addressed in out work.

One may well ask the following: We know that in classical mechanics, such constraints only reduce the number of degrees of freedom, and we also know that some quantum systems are already treated as two or one dimensional (like a two dimensional quantum gas, for example); so why should we expect to obtain a different picture than we have already? Although we cannot give any experimental evidence in our work, it is revealed that some interesting phenomena will occur under the influence of a curved lattice. There are a number of interesting studies on this issue, which assert that we should expect small quantum mechanical effects depending on geometry, see for example \cite{Ferrari, Ogawa}. What is more interesting is that, people have obtained contradicting results \cite{Ogawatek} from different descriptions of the same system. Since one can formulate a well defined physical problem, which can be realized at least in principle, there should be one unique correct result involving possible geometrical effects. Up to now, people tried to understand the differences between two methods which are most often addressed, and see in what circumstances those different approaches gave the same result \cite{Ikegami, Golovnev}. 

In our work, we will try to explain those two approaches and re-derive some of the results present in the literature (see \cite{Ferrari, Ogawa, Ikegami, Burgess}). Firstly, some necessary information about geometry will be given. Then, following the discussion in \cite{Ferrari}, we will analyze the first approach which can be classified as a geometrical approach. This treatment relies on the geometrical relations which are valid in the three dimensional Euclidean space, and begins with writing the relevant quantum mechanical equation of motion, which will be the well known Schr\"odinger equation at the first place. It is of course possible to consider the problem in $N$ dimensional space, but for the sake of simplicity we will not deal with the general case. Next, we will explain a quantization procedure which has first been proposed by Dirac. As one can immediately understand, the discussion will begin from classical mechanics, and we  will try to obtain proper quantum mechanical relations (if there are) via Dirac's quantization procedure. This procedure is already formulated for a generic number of dimensions, so our discussion will involve an $N$ dimensional Euclidean space and we will try to find the relevant Schr\"odinger equation again. Then, we will compare the results. As the last point, we will try to apply the first approach to Dirac equation, which can be considered as the beginning of a discussion for relativistic systems involving fermions.  

\section{GEOMETRICAL PRELIMINARIES}
\label{sec:Geometrical Preliminaries}

\subsection{Curves in $E^{3}$}
In the next chapter, we will try to expand Schr\"odinger equation in powers of a coordinate in $E^{3}$, and try to decompose the equation into two equations; one involving dynamics along a surface on which a particle is supposed to be constrained, and another equation which involves the direction that is transverse to the surface. This task requires the knowledge of some properties of curves and extrinsic and intrinsic geometries of surfaces in Euclidean space. Furthermore, in order to interpret the results of Dirac's quantization procedure, which has also been mentioned in the introduction, one needs some information about the geometry of surfaces. For this reason, we will briefly mention some properties of curves and surfaces in $E^{3}$ relevant to our task, and give additional information when necessary. 

One can identify curves as \textquotedblleft  \textit{...paths of a point in motion. The rectangular (Cartesian) coordinates $(x,y,z)$ of the point can then be expressed as functions of a parameter $u$ inside a certain closed interval}\textquotedblright \cite{Struik}:
\begin{align}
x_{i}=x_{i}(u),\qquad i=1,2,3, \qquad u_{1}\leq u\leq u_{2}.
\end{align}

This representation is known as the analytic (or parametric in some texts) representation \cite{Struik}.

For the requirements of our discussion, we will assume that the parameter $u$ is real and the rectangular coordinates $x_{i}$ under consideration are real functions of $u$. 

The following integral gives the arc length of the curve between two points on it as a function of the parameter $u$ \cite{Struik}:
\begin{align}
s(u)=\int _{u_{0}}^{u}\sqrt{\sum _{i}(\dfrac{dx_{i}}{du'})^{2}}\; du'=\int _{u_{0}}^{u}\sqrt{\dfrac{d\mathbf{x}} {du'}\cdot \dfrac{d\mathbf{x}} {du'}}\; du'.
\end{align}  
We assume here that $\dfrac{d\mathbf{x}} {du}$ is never zero within the given interval. With this assumption, one may parametrize the curve equally well with the arc length $s$. When this is the case, the vector $\mathbf{t}\equiv d\mathbf{x}/ds$ is the unit tangent vector of the curve \cite{Struik}. There are two other important vectors attached to each point on a curve. The first one is the principal normal vector, defined via the equation $d\mathbf{t}/ds\equiv \kappa \mathbf{n}$, and the second one is the binormal vector defined as $\mathbf{b}\equiv \mathbf{t}\times \mathbf{n}$. These vectors satisfy the well known Serret-Frenet formulae \cite{Mitchell} (in some texts, the formulae are included as the formulae of Frenet, see for example \cite{Struik}):
\begin{eqnarray}
\begin{bmatrix} 
d\mathbf{t}/ds \\ 
d\mathbf{n}/ds\\ 
d\mathbf{b}/ds  
\end{bmatrix}  = \begin{bmatrix}
0 & \kappa  & 0 \\ 
-\kappa & 0 & \tau  \\ 
0 & -\tau  & 0 
\end{bmatrix}\times \begin{bmatrix}
\mathbf{t} \\ 
\mathbf{n} \\ 
\mathbf{b} 
\end{bmatrix} 
\end{eqnarray}  
where $\kappa $ and $\tau $ are the curvature and torsion of the curve, respectively. 

\subsection{Surfaces in $E^{3}$}
There are various ways of representing a surface in the 3 dimensional Euclidean space ($E^{3}$). One is the familiar representation in which the surface is defined via an equation like $ f(x,y,z)=0 $, where $x, y$ and $z$ are the Cartesian coordinates and $f$ is a scalar function of those coordinates. One can also represent a surface with the Cartesian coordinates $x, y, z$ being real functions of two independent real parameters $u$ and $v$, called the \textquoteleft analytic (parametric) representation\textquoteright  \textit{in a certain closed interval} \cite{Struik}: 
\begin{align}
x_{i}=x_{i}(u,v), \qquad u_{1}\leq u \leq u_{2}, \qquad v_{1}\leq v \leq v_{2}.
\label{1.1} 
\end{align}
Here, $i=1,2,3$ stand for $x,y,z$, respectively. 

It is generally assumed that the Cartesian coordinates are differentiable functions of the parameters up to a sufficient order. In fact, for all practical purposes, one may assume here that they are differentiable to all orders. This way one may expand $x_{i}$ around a certain point $(u_{0},v_{0})$ \cite{Struik}:

\begin{align}
x_{i}(h,k) &=x_{i}(u_{0},v_{0}) +h\left( \dfrac{\partial x_{i}}{\partial u} \right)_{0}+k\left(\dfrac{\partial x_{i}}{\partial v} \right)_{0}
+\dfrac{1}{2!}\left[\left(h\dfrac{\partial}{\partial u} + k\dfrac{\partial}{\partial v}\right)^{2} x_{i} \right]_{0} 
+\cdots \nonumber \\ 
& +\dfrac{1}{n!}\left[\left(h\dfrac{\partial}{\partial u} + k\dfrac{\partial}{\partial v}\right)^{n} x_{i} \right]_{0}+\cdots . 
\end{align}
One also requires the parameters $u$ and $v$ to be independent so that the derivatives are independent. For this reason, the coordinate transformation matrix \cite{Struik}
\begin{align}
M\equiv \dfrac{\partial (x,y,z)}{\partial (u,v)},
\end{align}
must have rank 2. Notice that we are dealing with a general coordinate transformation here. 

There are two important objects which are now to be defined. 

\subsubsection{First Fundamental Form}
Consider some generic point $P$ which lies on a surface $\Sigma$ and let $\Sigma$ be parametrized by $u$ and $v$. Let $\mathbf{r}=\mathbf{r}$(u,v) be the position vector of $P$. If we now consider the squared distance between $P$ and some $P'$ which is in the neighborhood of $P$, we write:
\begin{align}
ds^{2} &=d\mathbf{r}\cdot d\mathbf{r}=\left(\dfrac{\partial \mathbf{r}(u,v)}{\partial u}du+\dfrac{\partial \mathbf{r}(u,v)}{\partial v}dv \right)\cdot \left(\dfrac{\partial \mathbf{r}(u,v)}{\partial u}du+\dfrac{\partial \mathbf{r}(u,v)}{\partial v}dv \right) \nonumber\\ 
&\equiv g_{\mu \nu}(q)dq^{\mu}dq^{\nu}, \qquad \mu ,\nu =1,2 \label{I},
\end{align}
where we identify $u,v$ with $q^{1},q^{2}$ respectively. $g_{\mu \nu}$ is the metric tensor of $\Sigma$, and the expression for $ds^{2}$ is called the first fundamental form of $\Sigma$ \cite{Struik}. Here we also use the Einstein summation convention, which means that when an index appears twice in an expression there is summation over it. Equation \ref{I} completely determines the intrinsic geometry of the surface $\Sigma $. 

\subsubsection{Second Fundamental Form}
One immediately notices that linear combinations of the vectors $\dfrac{\partial \mathbf{r}}{\partial u}$ and $\dfrac{\partial \mathbf{r}}{\partial v}$ can be used for expressing all vectors tangent to $\Sigma$ at $P$, so they naturally form a basis for those vectors. This means, the plane tangent to $\Sigma$ at $P$, which is the tangent plane of $\Sigma$ at $P$ \cite{Struik}, is spanned by these vectors. When a vector perpendicular to $\Sigma$ at $P$ is added to the naturally arising basis under consideration, one obtains an alternative basis for $E^{3}$. This way, at least the points in the immediate neighborhood of the surface $\Sigma$ can be expressed using this basis. We are going to discuss this issue later. 

The vector perpendicular to $\Sigma$ at $P$ which is of unit length and parallel to the vector $\nabla f(x,y,z)$ (where $f(x,y,z)=0$ defines $\Sigma$) is the unit normal vector $\mathbf{N}(u,v)$ of $\Sigma$ at $P$. Let us now consider the change in this vector as it moves from $P$ to $P'$ along some curve on $\Sigma$:
\begin{align}
d\mathbf{N}(u,v)=\dfrac{\partial \mathbf{N}(u,v)}{\partial u}du+\dfrac{\partial \mathbf{N}(u,v)}{\partial v}dv.   
\end{align}

Notice that $d\mathbf{N}(u,v)$ is tangent to $\Sigma$, just as $d\mathbf{r}(u,v)$ is. The second fundamental form is now defined to be \cite{Struik}:
\begin{align}
d\mathbf{N}(u,v)\cdot d\mathbf{r}(u,v)\equiv H_{\mu \nu}(q)dq^{\mu}dq^{\nu} \label{II}
\end{align}

[In \cite{Struik}, there is an overall minus sign in the definition, which does not affect the result included in this work.] The equation \ref{II} determines the extrinsic geometry of the surface $\Sigma $. Here, $H_{\mu \nu}$ is a symmetric matrix. Since both $d\mathbf{N}(u,v)$ and $d\mathbf{r}(u,v)$ are tangent to $\Sigma$, one may express $d\mathbf{N}(u,v)$ in terms of $\dfrac{\partial \mathbf{r}(u,v)}{\partial u}$ and $\dfrac{\partial \mathbf{r}(u,v)}{\partial v}$. Remembering that we have identified $u,v$ as $q^{1},q^{2}$:
\begin{align}
\dfrac{\partial \mathbf{N}(q)}{\partial q^{\mu}}\equiv \alpha _{\mu}~^{\nu} \; \dfrac{\partial \mathbf{r}(q)}{\partial q^{\nu}} \label{alpha}
\end{align}
where the two equations for the two components of $\dfrac{\partial \mathbf{N}(q)}{\partial q^{\mu}}$ ($\mu =1,2$) are known as Weingarten equations \cite{Struik, daCosta}. In some texts in the literature, the matrix $\alpha$ is called Weingarten curvature matrix \cite{Ferrari}, or the extrinsic curvature \cite{Ikegami}.

There are a number of nice properties of the matrix $\alpha$ \cite{Ferrari}: 
\begin{align}
& M = \dfrac{1}{2}tr(\alpha), \nonumber\\
& A = det(\alpha). \\
& H_{\mu \nu}=\dfrac{1}{2}(\alpha _{\mu}~^{\lambda}g_{\lambda \nu}+g_{\mu \lambda}\alpha ^{\lambda}~_{\nu}), \label{I-II}
\end{align}
where $M$ is the mean curvature and $A$ is the Gaussian curvature of $\Sigma $. These properties will be used in the calculations in the following chapters.

\section{THIN LAYER METHOD}
\label{sec:Thin Layer Method}

\subsection{Treatment of Ferrari and Cuoghi}
The name thin layer method for constraining the equation of motion of a quantum mechanical particle to some surface is due to Golovnev \cite{Golovnev}. The method has been used by da Costa \cite{daCosta} for a particle under the influence of a constraining potential but free to move on a surface, and by Ferrari and Cuoghi \cite{Ferrari} for a particle under the influence of an external electromagnetic field, which is also constrained to move on a surface. Both treatments involve the relevant Schr\"{o}dinger equation ignoring the spin of the particle, and the constraining procedure begins at the level of the Schr\"{o}dinger equation, unlike Dirac's procedure, which will be discussed later. 

Since the treatment of Ferrari and Cuoghi (given in \cite{Ferrari}) is more general, it is useful to summarize this procedure here.

We begin by making a general coordinate transformation in $E^{3}$, and writing the metric with our new coordinates in the vicinity of some surface $\Sigma$. Let us assume that two of our new coordinates $q^{1},q^{2}$ correspond to the parameters $u,v$ of the previous section, and $q^{3}$ be defined as the distance from $\Sigma$. Assuming that all $q^{\mu}, \; \mu =1,2,3$ are independent, we have three coordinates which define points uniquely at least in the vicinity of $\Sigma$. Let $\mathbf{r}(q^{1},q^{2})$ be the position vector of a generic point $P$ lying on $\Sigma$ and $\mathbf{N}(q^{1},q^{2})$ be the unit normal vector of the surface at $P$. Then, the position vector of some point $Q$ lying close to $\Sigma$ and a distance $q^{3}$ away from $P$ is $\mathbf{R}(q^{1},q^{2},q^{3})=\mathbf{r}(q^{1},q^{2}) + q^{3}\mathbf{N}(q^{1},q^{2})$. Then we have the following metric:
\begin{align}
G_{\mu \nu} &=\dfrac{\partial \mathbf{R}} {\partial q^{\mu}}\cdot \dfrac{\partial \mathbf{R}} {\partial q^{\nu}} \nonumber \\
G_{i j} &=\left(\dfrac{\partial \mathbf{r}} {\partial q^{i}}+q^{3}\dfrac{\partial \mathbf{N}} {\partial q^{i}} \right)\cdot \left(\dfrac{\partial \mathbf{r}} {\partial q^{j}}+q^{3}\dfrac{\partial \mathbf{N}} {\partial q^{j}} \right), \qquad i,j=1,2 \nonumber \\
G_{i 3} &=0, \qquad G_{3 3}=1, 
\end{align}   
where the scalar product of the vectors is taken with respect to the Euclidean metric $diag(1,1,1)$. Now, notice that the Weingarten matrix $\alpha$ enters into the expression for $G_{ij}$:
\begin{align}
G_{ij} &= \dfrac{\partial \mathbf{r}} {\partial q^{i}}\cdot \dfrac{\partial \mathbf{r}} {\partial q^{j}} + q^{3}\left(\dfrac{\partial \mathbf{r}} {\partial q^{i}}\cdot \alpha _{j}~^{k}\dfrac{\partial \mathbf{r}} {\partial q^{k}} + \alpha ^{k}~_{i}\dfrac{\partial \mathbf{r}} {\partial q^{k}}\cdot \dfrac{\partial \mathbf{r}} {\partial q^{j}}\right) + (q^{3})^{2}\alpha _{i}~^{k}\alpha ^{l}~_{j}\dfrac{\partial \mathbf{r}} {\partial q^{k}}\cdot \dfrac{\partial \mathbf{r}} {\partial q^{l}} \nonumber \\
&=g_{ij}+2q^{3}H _{ij}+(q^{3})^{2}\alpha _{i}~^{k} g_{kl}\alpha ^{l}~_{j}, \qquad i,j=1,2,
\label{metricdacosta}
\end{align}
where we used the definitions of the first and second fundamental forms and the relation between them (equations \ref{I}, \ref{II}, \ref{I-II}). Here one may argue that this expression for the metric tensor is at $O((q^{3})^{2})$, while the expression for the position vector of $Q$ is at $O(q^{3})$. However, we did not begin with the Taylor expansion for the position vector of $Q$, which is also possible. So, with our choice of the curvilinear coordinates $q^{\mu}$, the expression for $\mathbf{R}$ is exact. One may still expand $\mathbf{R}$ in powers of $q^{3}$, but the metric tensor obtained by da Costa \cite{daCosta} and Ferrari and Cuoghi \cite{Ferrari} will be a special case then. We will see later that beginning with the assumption that the third coordinate is orthogonal to our surface will bring some conditions on the terms appearing in the expansion. 

Now we are ready to write the Schr\"{o}dinger equation. Let $\mathbf{A}(q)$ be the vector potential and $\Phi(q)$ be the scalar potential corresponding to our external field. Let us define the gauge covariant derivative \cite{Ferrari} as $D_{\nu}\equiv \nabla_{\nu} - \dfrac{iQ}{\hbar}A_{\nu}$, where Q is the charge of our particle. We also define \textit{a gauge covariant derivative for the time variable} \cite{Ferrari} as $D_{0}\equiv \partial _{t} + \dfrac{iQ}{\hbar}\Phi$. Using the well known expression $\nabla ^{2}\Psi=\dfrac{1}{\sqrt{G}}\partial _{\mu}(\sqrt{G} G^{\mu \nu} \partial _{\nu} \Psi)$ for the Laplacian of the function $\Psi$, where $G\equiv det(G_{\mu \nu})$, we obtain, by direct substitution, the following equation \cite{Ferrari}:
\begin{align}
i\hbar D_{0}\Psi = \dfrac{1}{2m}\Bigg[-\dfrac{\hbar ^{2}}{\sqrt{G}}\partial _{\mu}(\sqrt{G}G^{\mu \nu}\partial _{\nu}\Psi) + \dfrac{iQ\hbar}{\sqrt{G}}\partial _{\mu}(\sqrt{G}G^{\mu \nu}A_{\nu})\Psi + 2iQ\hbar G^{\mu \nu}A_{\nu}\partial _{\mu}\Psi + Q^{2}G^{\mu \nu}A_{\mu}A_{\nu}\Psi \Bigg].
\label{schrodinger}
\end{align}

First, notice that by direct calculation:
\begin{align}
\sqrt{G}=\sqrt{g}\left (1+q^{3}Tr(\alpha)+(q^{3})^{2}det(\alpha)\right ),
\end{align}
where $Tr(\alpha )=\alpha _{i}~^{i}$ and $det(\alpha )=det(\alpha _{i}~^{j})$.

Now, let us consider the normalization integral:
\begin{align}
\int d^{3}x\Psi ^{*}(x)\Psi(x) &=\int d^{3}q\sqrt{G}\Psi ^{*}(q)\Psi(q) \nonumber \\
&=\int d^{3}q\sqrt{g}\left (1+q^{3}Tr(\alpha)+(q^{3})^{2}det(\alpha)\right ) \Psi ^{*}(q)\Psi(q)=1
\end{align}

If we now define a new wave function $ \chi (q) $ such that \cite{Ferrari}:
\begin{align}
\chi (q) = \Psi (q) \left (1+q^{3}Tr(\alpha)+(q^{3})^{2}det(\alpha)\right )^{1/2},
\end{align}
the normalization integral becomes:
\begin{align}
\int d^{3}q\sqrt{g}\left (1+q^{3}Tr(\alpha)+(q^{3})^{2}det(\alpha)\right )\Psi ^{*}(q)\Psi(q)=\int d^{2}qdq^{3}\sqrt{g}\chi ^{*}(q)\chi(q).
\end{align}

Notice the difference in the integral measure. This way, we also separate the normalization integral into two parts, which we will now try to do for the Schr\"{o}dinger equation, using the new wave function $\chi (q)$. Introducing an extra constraining potential $V_{\lambda}(q^{3})$ and calculating the derivatives of $\left (1+q^{3}Tr(\alpha)+(q^{3})^{2}det(\alpha)\right )^{1/2}$, we get the following equation in the limit $q^{3}\rightarrow 0$ \cite{Ferrari}:
\begin{align}
i\hbar D_{0}\chi & =\dfrac{1}{2m} \Bigg[-\dfrac{\hbar ^{2}} {\sqrt{g}}\partial _{i}(\sqrt{g}g^{ij}\partial _{j}\chi)+\dfrac{iQ\hbar}{\sqrt{g}}(\sqrt{g}g^{ij}A_{j})\chi +2iQ\hbar g^{ij}A_{i}\partial _{j}\chi +Q^{2}(g^{ij}A_{i}A_{j}+(A_{3})^{2})\chi \nonumber\\
& -\hbar ^{2}(\partial _{3})^{2}\chi +iQ\hbar (\partial _{3}A_{3})\chi +2iQ\hbar A_{3}(\partial _{3}\chi )-\hbar ^{2} \left(\left(\dfrac{1}{2}Tr(\alpha)\right)^{2}-det(\alpha) \right)\chi \Bigg] + V_{\lambda}(q^{3})\chi,
\end{align}
noticing that
\begin{align}
&\lim _{q^{3}\to\ 0}\left (1+q^{3}Tr(\alpha)+(q^{3})^{2}det(\alpha)\right )^{\pm 1/2} =1, \nonumber \\
&\lim _{q^{3}\to\ 0}\partial _{3}\left ((1+q^{3}Tr(\alpha)+(q^{3})^{2}det(\alpha)\right )^{\pm 1/2}) =\pm Tr(\alpha)/2, \nonumber \\
&\lim _{q^{3}\to\ 0}\partial _{3}^{2} \left((1+q^{3}Tr(\alpha)+(q^{3})^{2}det(\alpha)\right )^{\pm 1/2}) =\pm det(\alpha), \nonumber \\  
&\lim _{q^{3}\to\ 0}\partial _{i}\left ((1+q^{3}Tr(\alpha)+(q^{3})^{2}det(\alpha)\right )^{\pm 1/2}) =0.
\end{align}

Ferrari and Cuoghi call the term $V_{s}\equiv \dfrac{-\hbar ^{2}}{2m} \left(\left(\dfrac{1}{2}Tr(\alpha)\right)^{2}-det(\alpha) \right)$ as the \textquoteleft geometric potential\textquoteright . In order to get rid of the terms that indicate the coupling of the surface coordinates and the coordinate normal to the surface, one asks if a gauge transformation is possible. Ferrari and Cuoghi discuss that the equation (2.3) is gauge invariant (see Appendix A), so one can choose a gauge in which $A_{3}=0$ using the function $\gamma =-\int _{0}^{q^{3}} A_{3}(q^{1},q^{2},z)dz$ \cite{Ferrari} for the gauge transformation. Then, the equation (2.8) can be split into two equations, while the wave function is now written as $\chi (q)=\chi _{n}(q^{3})\chi _{s}(q^{1},q^{2})$ \cite{Ferrari}:
\begin{align}
i\hbar D_{0}\chi _{n}& = -\dfrac{\hbar ^{2}}{2m}(\partial _{3})^{2}\chi _{n}+V_{\lambda}(q^{3})\chi _{n}, \\
i\hbar D_{0}\chi _{s}& = \dfrac{1}{2m} \Bigg[-\dfrac{\hbar ^{2}} {\sqrt{g}}\partial _{i}(\sqrt{g}g^{ij}\partial _{j}\chi _{s})+\dfrac{iQ\hbar}{\sqrt{g}}(\sqrt{g}g^{ij}A_{j})\chi _{s}+2iQ\hbar g^{ij}A_{i}\partial _{j}\chi _{s} +Q^{2}g^{ij}A_{i}A_{j}\chi _{s}\Bigg]+V_{s}\chi _{s}. 
\end{align}          

Having decoupled the dynamics on the surface and in the normal direction, one may now treat the above equations separately to solve for the wave function $\chi $, and calculate the expectation value of any relevant physical quantity on the surface without referring to the \textquoteleft external world\textquoteright, that is, using only the quantities defined on $\Sigma$ being independent of the normal direction.

\subsection{Taylor Expanding the Position Vector}
As we noted, the squeezed coordinate $q^{3}$ is chosen to be the distance from the surface to which the particle is constrained in the analyses of da Costa and Ferrari and Cuoghi. Another possible approach would be to consider a more general case of a coordinate transformation in which two of the coordinates again parametrize the surface and the third one is again chosen to be orthogonal to the surface. In Euclidean space, the distance from the surface coincides with the third coordinate of the curvilinear set if and only if the coordinate  curves of $q^{3}$ are straight lines. This can be understood by noticing the fact that in Euclidean space, the shortest path between two points is a straight line and the distance between the points is the length of that path. Then, when this is the case, the discussion simply reduces to the one given by da Costa and Ferrari and Cuoghi. One may also ask whether the coordinate $q^{3}$ corresponds to the arc length along the coordinate curve, or not. Let us first consider the general case, then answer this question.  

In general, assuming that $q^{3}$ is orthogonal to the surface $\Sigma$, one may expand the position vector of a point in the neighborhood of $\Sigma$ as follows: 
\begin{align}
\mathbf{R}=\mathbf{R}_{0}+(\partial _{3}\mathbf{R})_{0}q^{3}+\dfrac{1}{2!}(\partial ^{2}_{3} \mathbf{R})_{0}(q^{3})^{2}+\dfrac{1}{3!}(\partial ^{3}_{3} \mathbf{R})_{0}(q^{3})^{3}+\cdots,
\end{align}
where the subscript $0$ implies that the quantity is evaluated at $q^{3}=0$. Notice that the first derivative of the position vector with respect to $q^{3}$ is always perpendicular to $\Sigma$, since it measures the rate of change of the position vector along a coordinate curve which is by assumption orthogonal to $\Sigma$. This expansion still handles all three dimensions, so the coordinates are by definition independent from each other. This fact will be useful soon. We have, for the metric tensor $G_{\mu \nu}=\partial _{\mu}\mathbf{R}\cdot \partial _{\nu}\mathbf{R}$, the usual definition. Now, our choice (or assumption) that the third coordinate is orthogonal to $\Sigma$, combined with the trivial fact just mentioned, brings the following conditions when one writes down the metric up to second order by direct calculation:
\begin{align}
& O(q^{3}): \quad \partial _{i}(\mathbf{R}_{0})\cdot \mathbf{R}''_{0}+\partial _{i}((\mathbf{R}')_{0})\cdot (\mathbf{R}')_{0}=0, \\
& O((q^{3})^{2}): \quad 2\partial _{i}(\mathbf{R}'_{0})\cdot \mathbf{R}''_{0} + \partial _{i}(\mathbf{R}_{0})\cdot \mathbf{R}'''_{0} + \partial _{i}(\mathbf{R}''_{0})\cdot \mathbf{R}'_{0}=0, 
\end{align}
where the primes denote differentiation with respect to $q^{3}$. At this point, let us assume that the third coordinate is chosen to be the arc length along the coordinate curve. Then, using the Serret-Frenet formulae mentioned in Chapter 1 Section 1, one has:
\begin{align}
\mathbf{R}=\mathbf{R}_{0}+\mathbf{t}_{0}q^{3}+\dfrac{1}{2!}(\kappa \mathbf{n})_{0}(q^{3})^{2}+\dfrac{1}{3!}(-\kappa \mathbf{t}+\tau \mathbf{b})_{0}(q^{3})^{3}+\cdots
\end{align}   
and for the above condition at $O(q^{3})$:
\begin{align}
\partial _{i}(\mathbf{R}_{0})\cdot (\kappa \mathbf{n})_{0}+\partial _{i}(\mathbf{t})_{0}\cdot \mathbf{t}_{0}=0
\end{align}
Now, notice that the second term is zero by definition of $\mathbf{t}$. So, the first term should also be zero. Now notice that, for nonzero $\kappa \mathbf{n}$, this is impossible, since by definition $\mathbf{n}$ is perpendicular to $\mathbf{t}$ and hence $\mathbf{n}_{0}$ is tangent to $\Sigma$. But a vector tangent to $\Sigma$ cannot be perpendicular to both of the vectors spanning $\Sigma$ at the same time, which is required by the condition. The only possibility is that $(\kappa \mathbf{n})_{0}=0$. This may somehow be satisfied locally on certain surfaces, but if we begin with a coordinate transformation which is globally valid (or valid at least within a finite region of space), then $\kappa \mathbf{n}=0$ should hold globally (or within the finite region considered). This means, one requires the coordinate curve of $q^{3}$ to be a straight line at least within a relevant finite region, like the neighborhood of $\Sigma$ under consideration which is regarded as a finite region of space at the beginning (before squeezing the system to the surface). Conversely, if the coordinate curve of $q^{3}$ is a straight line, then $q^{3}=a\lambda +b$, where $\lambda$ is the arc length parameter (and so the distance from $\Sigma$) and $a$ and $b$ are constants. This  result reduces our problem to the case considered by da Costa and Ferrari and Cuoghi. 

Now, we may concentrate on the other possible situation, in which $q^{3}$ is not the arc length along its coordinate curve. The metric tensor in this case becomes: 
\begin{align}
G_{ij} &=\partial _{i}\Bigg[\mathbf{R}_{0}+\mathbf{R'}_{0}q^{3}+\dfrac{1}{2!}\mathbf{R''}_{0}(q^{3})^{2}\Bigg]\cdot \partial _{j}\Bigg[\mathbf{R}_{0}+\mathbf{R'}_{0}q^{3}+\dfrac{1}{2!}\mathbf{R''}_{0}(q^{3})^{2}\Bigg ] \nonumber \\
&=\partial _{i}(\mathbf{R}_{0})\cdot \partial _{j}(\mathbf{R}_{0})+q^{3}\left[\partial _{i}(\mathbf{R}_{0})\cdot \partial _{j}(\mathbf{R'}_{0})+\partial _{i}(\mathbf{R'}_{0})\cdot \partial _{j}(\mathbf{R}_{0})\right ]\nonumber \\
&+\dfrac{1}{2}(q^{3})^{2}\left[\partial _{i}(\mathbf{R}_{0})\cdot \partial _{j}(\mathbf{R''}_{0})+\partial _{i}(\mathbf{R''}_{0})\cdot \partial _{j}(\mathbf{R}_{0})+2\partial _{i}(\mathbf{R'}_{0})\cdot \partial _{j}(\mathbf{R'}_{0})\right],\nonumber \\
G_{33} &=\mathbf{R'}_{0}\cdot \mathbf{R'}_{0}+2q^{3}\mathbf{R'}_{0} \cdot \mathbf{R''}_{0}+(q^{3})^{2}\left[\mathbf{R''}_{0}\cdot \mathbf{R''}_{0}+\mathbf{R'}_{0}\cdot \mathbf{R'''}_{0}\right],
\end{align}
up to $O((q^{3})^{2})$. Notice that the $O(q^{3})$ term of $G_{ij}$ is the same with the one obtained in the previous section, since $\partial _{i}\mathbf{R}_{0}$ is tangent to $\Sigma$ and so $\partial _{i}(\mathbf{R}_{0})\cdot \partial _{j}(\mathbf{R'}_{0})=\partial _{i}(\mathbf{R}_{0})\cdot \partial _{j}(\mathbf{t}_{0})=\partial _{i}(\mathbf{R}_{0})\cdot \partial _{j}(\mathbf{N})$. But there are additional terms in the second order term, and also $G_{33}\neq 1$. One may still handle the problem by calculating the contributions to the determinant of $G_{\mu \nu}$, but the measure of integrations which arise in the normalization and in the calculations for expectation values also changes, and takes the form:
\begin{align}
d^{3}q=dq^{3}d^{2}q\sqrt{G_{33}}\sqrt{g_{new}}
\end{align}
Although one may still find the expansion in powers of $q^{3}$ corresponding to $\sqrt{g_{new}}$, the factor $\sqrt{G_{33}}$ spoils the separation of the integral into a surface integral and an integral along the normal direction which are independent of each other. This means, it is not guaranteed that we still have the chance for calculating expectation values or carry out normalization on the surface without any reference to the \textquoteleft external world\textquoteright . 

\subsection{Spin-Magnetic Field Interaction}
It would be interesting to consider the spin-magnetic field interaction within the context of the thin layer method. Let us consider the following interaction
\begin{align}
\widehat{H}_{I}=-\dfrac{e}{2mc}\mathbf{B}\cdot \mathbf{S}
\end{align}
where $e$ is the magnitude of the electron charge, $m$ is the electron mass, $c$ is the speed of light in vacuum and $\mathbf{S}$ is the $2\times 2$ spin operator composed of the well known Pauli spin matrices. In terms of a vector potential $\mathbf{A}$, this term can be written as:
\begin{align}
\widehat{H}_{I}=-\dfrac{e}{2mc}\nabla \times \mathbf{A}\cdot \mathbf{S}=-\dfrac{e}{2mc}\epsilon ^{\mu \nu \lambda}(\nabla _{\mu}A_{\nu})S_{\lambda}
\label{inth}
\end{align}
where $\epsilon ^{123}=1$. In order to calculate the effects of the curvature of $\Sigma$, we need to calculate the connection coefficients in the coordinates used. Using the well known coordinate expression for the coefficients:
\begin{align}
\Gamma ^{\mu}_{\nu \lambda}=\dfrac{1}{2}G^{\mu \rho}(G_{\rho \nu ,\lambda}+G_{\rho \lambda ,\nu}-G_{\nu \lambda ,\rho}),
\end{align}
and the inverse of $G_{\mu \nu}$ to $O(q^{3})$:
\begin{align}
G^{ij}=g^{ij}-2q^{3}H^{ij},
\end{align}
we get the following expressions:
\begin{align}
\Gamma ^{3}_{\nu 3} &=\Gamma ^{\mu}_{33}=0, \nonumber \\
\Gamma ^{3}_{ij} &=-H_{ij}-q^{3}\alpha _{i}~^{k}g_{kl}\alpha ^{l}~_{j}, \nonumber \\ 
\Gamma ^{k}_{i3} &=g^{kl}H_{li}-q^{3}H^{km}H_{mi},\nonumber \\
\Gamma ^{k}_{ij} &=\tilde{\Gamma} ^{k}_{ij}+q^{3}\Big[g^{kl}(H_{li,j}+H_{lj,i}-H_{ij,l})-H^{kl}(g_{li,j}+g_{lj,i}-g_{ij,l})\Big],
\label{connections}
\end{align}
where $\tilde{\Gamma} ^{k}_{ij}=\dfrac{1}{2}g^{kl}(g_{li,j}+g_{lj,i}-g_{ij,l})$. We calculated the connection coefficients up to $O(q^{3})$ since there are only first derivatives present in the expression considered. Although the expression for $G_{\mu \nu}$ is exact in the case analyzed by da Costa, the inverse is calculated to first order in $q^{3}$ since we will at the end take the limits of everything at $q^{3}=0$. Using these, we have the following expressions for $\nabla \times \mathbf{A}$:
\begin{align}
(\nabla _{i}A_{j})_{0}=(\partial _{i}A_{j})_{0} - \tilde{\Gamma} ^{k}_{ij}(A_{k})_{0}+H_{ij}(A_{3})_{0},\nonumber \\
(\nabla _{3}A_{j})_{0}= (\partial _{3}A_{j})_{0} - g^{kl}H_{lj}(A_{k})_{0},\nonumber \\
(\nabla _{j}A_{3})_{0}=(\partial _{j}A_{3})_{0} - g^{kl}H_{lj}(A_{k})_{0}.
\end{align}
So, finally, we get the following expression for the interaction term:
\begin{align}
\widehat{H}_{I}=-\dfrac{e}{2mc}[S_{3}B^{3}+\epsilon ^{ij3}S_{i}(\partial _{j}A_{3}-\partial _{3}A_{j})]_{0},
\end{align}
by putting the above terms directly into (\ref{inth}). The result that can be derived from this expression is that spin-magnetic field interaction always includes coupling with external world and this is unavoidable. Otherwise, the interaction term will simply be zero, meaning that there is no interaction at all.

\section{DIRAC'S QUANTIZATION PROCEDURE}
\label{sec:Dirac's Quantization Procedure}

\subsubsection{Overview of the Procedure}

Another method of handling geometrical constraints imposed on physical systems was discussed by Dirac in 1950s \cite{Dirac} (see also the references therein). The method involved modification of canonical quantization procedure beginning from the Lagrangian level, and carrying the constraints properly to the Hamiltonian formalism. Dirac also modified the terminology, and used the word \textquoteleft constraint\textquoteright \, in a more general manner. We will also follow this usage, and indicate the usual geometrical constraints explicitly where necessary. 

We will, in this section, briefly summarize the relevant parts of the first two chapters of \cite{Dirac}, which cover all issues relevant to the problem discussed in the previous chapters of our work. The discussion in \cite{Dirac} about this issue begins with a reminder of the ideas of Lagrangian formalism and transition to Hamiltonian formalism. Although the motivation is to formulate some Hamiltonian formalism for a field theory, Dirac begins the discussion with particle dynamics. 

As is well known, the classical equations of motion for a particle moving in an $N$ dimensional space can be obtained by extremizing the so-called action integral:
\begin{align}
S=\int L(q,\; \dot{q},\; t)dt, \nonumber \\
\delta S=0,
\end{align}  
where $q_{n}, \; (n=1,\cdots ,N)$ are the generalized coordinates of the particle, and the over dot denotes differentiation with respect to time $t$. Transition to the Hamiltonian formalism begins by defining the momentum variables conjugate to these coordinates as follows:
\begin{align}
p_{n}\equiv \dfrac{\partial L}{\partial \dot{q}_{n}}.
\end{align}
Usually, one assumes that the momenta are all independent from each other. Here, Dirac notes the following \cite{Dirac}: \textquotedblleft \textit{We want to allow for the possibility of these momenta not being independent functions of the velocities. In that case, there exist certain relations connecting the momentum variables, of the type $\phi (q,p) = 0$.}\textquotedblright 

This statement seems to be rather implicit, since one cannot directly visualize what sort of relations can appear while calculating the momenta. There is one easier way of revealing these relations via inserting the geometrical constraints which are usually stated as $f(q)=0$ as functions of coordinates into the Lagrangian with the motivation to use the method of Lagrange multipliers, as used by other people in the literature \cite{Ogawa, Ogawatek, Ikegami}:
\begin{align}
L_{new}=L+\lambda _{s}f_{s}(q),
\end{align}
where $f_{s}(q)=0$ are the functions defining the geometrical constraints and $\lambda _{s}$ are the corresponding Lagrange multipliers, which are also treated as dynamical variables in the method of Lagrange multipliers. We also adopt here the summation convention used in the previous chapters. One deals with this new Lagrangian and finds possible relations among the momenta. But of course, this is not necessarily the most general way of using Dirac's procedure, and as we will see, the results predicted by the procedure will depend on the explicit expressions of the geometrical constraints \cite{Ikegami}. 

Having noted this fact, we may return to Dirac's discussion. Dirac calls the relations $\phi _{m} (q,p)=0, \; m=1,\cdots ,M$ as \textquoteleft primary constraints of the Hamiltonian formalism\textquoteright \, \cite{Dirac}. Notice that they appear at the stage of defining the momenta. Then, one defines the Hamiltonian in the usual way:
\begin{align}
H\equiv p_{n}\dot{q}_{n}-L.
\end{align}
However, as is also well known, Hamiltonian defined in this way is not unique. One can equally well write the following Hamiltonian, simply adding some \textquoteleft zeros\textquoteright \, to the usual one \cite{Dirac}:
\begin{align}
H^{*} =H+c_{m}\phi _{m}.
\end{align}
As stated by Dirac, the coefficients $c_{m}$ can be functions of $q_{n}$ and $p_{n}$. The crucial point is the following \cite{Dirac}: \textquotedblleft \textit{$H^{*}$ is then just as good as $H$; our theory cannot distinguish between $H$ and $H^{*}$. The Hamiltonian is not uniquely determined.}\textquotedblright
 
Remembering that the Hamiltonian equations of motion are 
\begin{align}
\dot{q}_{n}=\dfrac{\partial H}{\partial p_{n}}, \quad \dot{p}_{n}=-\dfrac{\partial H}{\partial q_{n}},
\end{align}
they now become \cite{Dirac}
\begin{align}
\dot{q}_{n}=\dfrac{\partial H}{\partial p_{n}}+u_{m}\dfrac{\partial \phi _{m}}{\partial p_{n}}, \quad \dot{p}_{n}=-\dfrac{\partial H}{\partial q_{n}}-u_{m}\dfrac{\partial \phi _{m}}{\partial q_{n}},
\label{dotlar}
\end{align} 
as one decides to use $H^{*} $ instead of $H$. Here, $u_{m}$ are unknown coefficients (we might have kept the $c_{m}$'s and so their derivatives, but here we follow Dirac). Although we will not use Hamilton's equations of motion, it is important to note how they are affected when certain constraints are imposed on the system. 

In order to proceed more consistently and in a way which is suitable for quantum mechanics, one needs to define Poisson brackets. As is well known, the Poisson bracket of a quantity with the Hamiltonian gives the time evolution of that quantity in the Hamiltonian formalism in classical mechanics. In the canonical quantization scheme which is already familiar to us, classical Poisson brackets are replaced by $-\dfrac{i}{\hbar }$ times the commutator of the entities involved, and also the functions representing physical entities are replaced by operators. The usual definition of Poisson brackets is the following:
\begin{align}
[f,g]\equiv \dfrac{\partial f}{\partial q_{n}}\dfrac{\partial g}{\partial p_{n}} -\dfrac{\partial f}{\partial p_{n}}\dfrac{\partial g}{\partial q_{n}}
\label{Poisdef}
\end{align}
and it satisfies the following important properties:
\begin{align}
&[f,g]=-[g,f], \label{Pois1}\\
&[f_{1}+f_{2},g]=[f_{1},g]+[f_{2},g], \label{Pois2}\\
&[f_{1}f_{2},g]=f_{1}[f_{2},g]+[f_{1},g]f_{2}, \label{Pois3}\\
&[f,[g,h]]+[g,[h,f]]+[h,[f,g]]=0 \; (Jacobi \; Identity). \label{Pois4}
\end{align}

Here Dirac makes an important remark. With the above definition of Poisson brackets, one cannot write the Poisson bracket of a quantity which may not be a function of $q_{n}$ and $p_{n}$, but may be a function of time \cite{Dirac}. This is the case for the unknown coefficients $u_{m}$ above. In order to avoid such a situation and express the time evolution of any quantity in some Hamiltonian formalism on the same footing with the time evolution of functions of $q_{n}$ and $p_{n}$, Dirac suggests to leave the usual definition of Poisson brackets and take the four equations (\ref{Pois1}), (\ref{Pois2}), (\ref{Pois3}), (\ref{Pois4}) stated above as the defining properties for Poisson brackets \cite{Dirac}. Now, notice that:
\begin{align}
\dot{g}=\dfrac{\partial g}{\partial q_{n}}\dot{q_{n}}+\dfrac{\partial g}{\partial p_{n}}\dot{p_{n}}=[g,H]+u_{m}[g,\phi _{m}],
\end{align}
for some $g=g(q,p)$ using the equations (\ref{dotlar}) and (\ref{Poisdef}). Here, Dirac asks whether one can write this equation as: 
\begin{align}
\dot{g}=[g,H+u_{m}\phi _{m}],
\end{align}
and gives the answer that, this is possible if we use the equations (\ref{Pois1})-(\ref{Pois4}) as the defining properties of the brackets. Then, we obtain:
\begin{align}
\dot{g}=[g,H+u_{m}\phi _{m}]=[g,H]+u_{m}[g,\phi _{m}]+[g,u_{m}]\phi _{m}.
\end{align}
Now, we remember that indeed $\phi _{m}=0$ and the last term should vanish. So it does. But why did we use this fact at this point rather than at the very beginning? Dirac's gives the answer and derives an important rule of quantizing constrained systems \cite{Dirac}: \textquotedblleft \textit{We have the constraints $\phi _{m}=0$, but we must not use one of these constraints \underline{before} working out a Poisson bracket. ... So we take it as a rule that Poisson brackets must all be worked out before we make use of the constraint equations.}\textquotedblright

As a reminder, he devises the following notation:
\begin{align}
\phi _{m}\approx 0
\end{align}
for the equations which are to be used after all Poisson brackets are worked out, and calls such equations as \textit{weak equations}; implying that the equations which are not of this sort are \textit{strong equations}. Then, the term $[g,u_{m}]\phi _{m}$ vanishes, and the expressions for $\dot{g}$ agree. 

From now on, the quantity $H+u_{m}\phi _{m}$ will be called the \textit{total Hamiltonian} \cite{Dirac} and will be denoted by $H_{T}$. 

At this point, it is necessary to state another important fact. The time evolution of the constraints should also be taken into account, and all time derivatives of the constraints should vanish to ensure consistency. For this reason, the equations stating that the time derivatives of the constraints vanish are called \textit{consistency conditions} \cite{Dirac}. An important question arises here: there can be an infinite number of consistency conditions, because the constraint functions can be differentiable to all orders; so which equations are the relevant consistency conditions to our problem?

Dirac gives the answer to this question by classifying the equations resulting from equating the time derivatives of the constraint functions to zero \cite{Dirac}:

\textbf{Case 0:} An inconsistency may occur. This implies that Lagrangian equations of motion are inconsistent. Dirac's example is the equations resulting from the Lagrangian $L=q$ which directly lead to $1=0$. So, the relevant cases are those in which Lagrangian equations of motion are consistent. 

\textbf{Case 1:} Some equations may reduce to $0=0$ with the help of primary constraints. 

\textbf{Case 2:} Some equations may reduce to equations which are independent of the coefficients $u_{m}$. They are then equations like $\chi =\chi (q,p)=0$, which means they are indeed new constraint equations giving new relations between the dynamical coordinates and their conjugate momenta. These are called \textit{secondary constraints}. For consistency, one should continue calculating the time derivatives of these functions to ensure consistency; that is, new consistency conditions may result from these functions.  

\textbf{Case 3:} The equation may reduce to one which brings conditions on $u_{m}$. At this point, the procedure for the relevant constraint (either primary or secondary) ends. 

So, the procedure up to the present level can be summarized as follows: write down the Lagrangian, then define the momenta and the Hamiltonian, determine the primary constraints, construct the total Hamiltonian, calculate the time derivatives of the constraints, classify them using the above classification. Case 1 equations bring nothing new, so they are trivial in a manner. Case 2 equations imply new constraints, so their time derivatives should be calculated and set equal to zero until Case 3 equations are reached. 

As Dirac himself notes \cite{Dirac}, \begin{quotation}
\textit{The secondary constraints will for many purposes be treated on the same footing as the primary constraints. ... They ought to be written as weak equations in the same way as primary constraints, as they are also equations which one must not make use of before one works out Poisson brackets.}
\end{quotation}

One may well ask the following question: Why do we terminate the procedure when Case 3 equations are reached? Dirac gives a detailed discussion, but it will be sufficient to give only the idea. Case 3 equations give $u_{m}$ as functions $U_{m}$ of $q_{n}$ and $p_{n}$ (because the Case 3 system of equations should have some solution if the equations of motion are consistent \cite{Dirac}). But, after working out the Poisson brackets, one may treat the constraint equations as strong equations; so, the constraint terms, together with their coefficients, will vanish. 

Of course, there will still be some arbitrariness in the Hamiltonian, because one may add terms which are weakly equal to zero with arbitrary coefficients to the solutions for $u_{m}$ obtained from Case 3 equations. Those arbitrary coefficients may be functions of time, and will be totally arbitrary. Dirac takes this as the reflection of some arbitrariness in the mathematical framework, like gauges in electrodynamics. Another interpretation might be that the number of independent degrees of freedom is less than the number of generalized dynamical coordinates, and there is still a freedom for observer choice. 

Here we should also define the first class and second class quantities. A quantity which is a function of $q_{n}$ and $p_{n}$ is labeled as first class if its Poisson bracket with all constraints vanishes at least weakly; and is second class otherwise \cite{Dirac}. It is noted in the text \cite{Dirac} that primary first class constraints are the generating functions of contact transformations, but it is not quite obvious that this fact can be made use of in the analysis. 

Now one is ready to define a bracket method for handling time evolution of quantities. For this, one should first notice that the total Hamiltonian is a first class quantity. Here, it is necessary to follow Dirac's reasoning directly to see why the total Hamiltonian is a first class quantity, and what this brings to us. 

First, remember that Case 3 equations can be solved for $u_{m}$ and then $u_{m}$ are obtained as functions of $q_{n}$ and $p_{n}$. Those Case 3 equations are indeed a system of non-homogeneous equations in the unknowns $u_{m}$. But since the system is non-homogeneous, one may add the solutions of a homogeneous system of equations like $V_{m}[\phi _{j},\phi _{m}]=0$. Here, $j=1,\cdots , J$ counts all constraints (primary and secondary). Then, if the independent solutions of $V_{m}[\phi _{j},\phi _{m}]=0$ are $V_{am}(q,p),\; a=1,\cdots ,A $, one may write $u_{m}(q,p)=U_{m}(q,p)+v_{a}(t)V_{am}(q,p)$. Notice that the coefficients $v_{a}$ can in general be functions of time, and are the totally arbitrary coefficients mentioned above. Then, the total Hamiltonian becomes \cite{Dirac}:
\begin{align}
H_{T}=H+U_{m}\phi _{m}+v_{a}V_{am}\phi _{m}\equiv H'+v_{a}\phi _{a}.
\end{align}
Notice that this total Hamiltonian is first class. $H'$ is first class since we impose $[\phi _{j},H']=0$ for consistency. $\phi _{a}=V_{am}\phi _{m}$ are first class due to involving the solutions of the homogeneous system of equations $V_{m}[\phi _{j},\phi _{m}]=0$. 

There is one more concept to be defined. It is what Dirac calls the \textquoteleft physical state\textquoteright \, of the system. In classical mechanics, one solves the Lagrangian equations of motion and calculates the time evolution of the system. This tells one that, what the observer should actually observe is described by the solutions of those equations. This gives the understanding about the physical state; the \textquoteleft physical state\textquoteright \, of the system concerns what the observers will actually observe. The solutions of the equations may seem to be different quantitatively due to observer choice, but a proper transformation from one observer to another one will reveal that different observers describe the same dynamics. Then, in Dirac's approach, the arbitrary features in the formalism should somehow not affect the state of the system, although they may change the values of the dynamical variables. So the following question arises: Which constraints do not affect the physical state of the system?

In classical mechanics, once the initial conditions of a system are completely known, all history of the system can be worked out. In other words, the initial physical state of the system determines the past and the future of the system completely. Now, consider some dynamical variable $g$, and let $g(t=0)=g_{0}$ be given. After a short time $\delta t$, $g$ evolves as \cite{Dirac}:
\begin{align}
g(\delta t)=g_{0}+\dot{g}\delta t=g_{0}+[g,H_{T}]\delta t=g_{0}+\delta t\left([g,H']+v_{a}[g,\phi _{a}]\right).
\end{align}
If we now change the set of $v_{a}$ values (making use of their arbitrariness), we will have a different $\dot{g}$. Taking the difference, we get:
\begin{align}
\Delta g(\delta t)=\delta t(v_{a}-v'_{a})[g,\phi _{a}]=\delta t\epsilon _{a}[g,\phi _{a}].
\end{align} 
Now, notice that $\phi _{a}$ are formed by primary constraints, and are also first class. If the physical state remains unchanged, as we require, we are led to the conclusion that primary first class constraints do not affect the physical state of the system. Also Dirac notes that the above equation means $\phi _{a}$ are generators of infinitesimal contact transformations, and further, conjectures that secondary first class constraints do not affect the physical state either. However, there are counterexamples of this conjecture, mentioned in the literature \cite{Hanneaux}. In any case, we will not deal with such examples and construct the procedure as if Dirac's conjecture is correct. This will amount to adding the secondary first class constraints to the Hamiltonian and defining the extended Hamiltonian \cite{Dirac}:
\begin{align}
H_{E}\equiv H_{T}+v'_{a'}\phi _{a'}.
\end{align}
Now, all independent accessible constraints are included in the Hamiltonian, and the time evolution of any system will be given as:
\begin{align}
\dot{g}=[g,H_{E}].
\end{align}
The last point in the construction of the theory is defining a proper bracket formalism. Then, the idea is simply converting the brackets into $-i/\hbar$ times commutators, like canonical quantization. 

In order not to deal with physically irrelevant degrees of freedom, Dirac defines a new bracket in terms of Poisson brackets \cite{Dirac}:
\begin{align}
[f,g]^{*}\equiv [f,g]-[f,\phi _{l}]\Delta ^{-1}_{kl}[\phi _{k},g], \\
\Delta _{kl}\equiv [\phi _{k},\phi _{l}].
\label{dbdef}
\end{align}
Here, $k,l=1,\cdots ,J$, count all constraints. These brackets satisfy the four important properties of Poisson brackets (\ref{Pois1})-(\ref{Pois4}) (see Appendix B for proofs). After defining these brackets, since all Poisson brackets have been worked out, one can treat the constraint equations as strong equations. This new bracket is often called the \textquoteleft Dirac bracket\textquoteright .

The issues mentioned in this section gave an outline of the relevant parts of Dirac's treatment in \cite{Dirac}. In the following sections, we will investigate various cases which have been discussed in the literature. 

\subsection{Treatment of Ogawa, Fujii and Kobushkin}
In this section, we will briefly summarize the discussion in \cite{Ogawa} and give some comments on the procedure. 

The discussion concerns an $N$ dimensional Euclidean space and begins from the Lagrangian level as Dirac's procedure requires. As we mentioned in the previous section, there is some arbitrariness in expressing the constraints within the formulation of the problem, and one option is to make use of the method of Lagrange multipliers; this has been preferred by Ogawa \textit{et. al}. The classical Lagrangian is then \cite{Ogawa}:
\begin{align}
L=\dfrac{1}{2}\dot{x}^{a}\dot{x}_{a}-V(x)+\lambda f(x),
\end{align}
where the particle is assumed to have unit mass for simplicity, $x^{a} \; (a=1, \cdots ,N)$ are the Cartesian coordinates of $E^{N}$ and $f(x)=0$ is the function expressing the geometrical constraint. As usual, the overdot denotes time derivative. In terms of the Cartesian coordinates, we have $p_{a}=\dfrac{\partial L}{\partial \dot{x}^{a}}=\dot{x}_{a}$ as momenta conjugate to $x^{a}$ and $p_{\lambda}=\dfrac{\partial L}{\partial \dot{\lambda}}\approx 0$ as our primary constraint \cite{Ogawa}. Following Dirac's procedure, one obtains (see Appendix B) \cite{Ogawa}:
\begin{align}
& H=\dfrac{1}{2}p_{a}p^{a}+V(x),\\
& [x^{a},p_{b}]_{D}=\delta ^{a}~_{b}-n^{a}n_{b},\nonumber \\
& [p_{a},p_{b}]_{D}=p^{c}(n_{b}\partial _{c}n_{a}-n_{a}\partial _{c}n_{b}),\nonumber \\
& [x^{a},x^{b}]_{D}=0,\\
& f(x)=0,\nonumber \\
& p^{a}\partial _{a}f=0,
\end{align}
where $n_{a}\equiv \dfrac{\partial f}{\partial x^{a}}$ is the unit normal vector of the $N-1$ dimensional hypersurface defined by $f(x)=0$. Notice that we changed our notation for Dirac brackets $[A,B]_{D}$, which was expressed as $[A,B]^{*}$ in the previous section. Ogawa \textit{et. al.} derive the classical equations of motion from Hamilton's equations using Dirac brackets and show that they are the expected equations of motion given by the usual Hamiltonian formalism (see Appendix B for details). 

At this point, a general coordinate transformation is performed \cite{Ogawa}:
\begin{align}
dx^{a}=\dfrac{\partial x^{a}}{\partial q^{\mu}}dq^{\mu},
\end{align}
where $\mu =0,\cdots ,N-1$. The authors choose the $q^{0}$ coordinate to be orthogonal to the hypersurface defined by $f(x)=0$. Then, $f(x)=0$ simply becomes $q^{0}=0$, and $p^{a}\partial _{a}f=\dot{x}^{a}\partial _{a}f=0$ becomes $\dot{x}^{a}\partial _{a}q^{0}=\dot{q}^{0}=0$. In terms of the new coordinates, we have the following metric:
\begin{align}
g_{\mu \nu}=\dfrac{\partial x^{a}}{\partial q^{\mu}}\dfrac{\partial x^{b}}{\partial q^{\nu}}\delta _{ab},
\end{align}
\begin{eqnarray}
[g_{\mu \nu}]=\begin{bmatrix}
g_{00} & 0 \\ 
 0 & g_{ij} 
\end{bmatrix}
\end{eqnarray}
Here, $g_{ij}$ corresponds to the first fundamental form of the hypersurface (i.e. the metric induced on the hypersurface). The classical equations of motion are again the expected equations. In terms of the new coordinates, the brackets are (see Appendix B for details):
\begin{align}
[q^{\mu},p_{\nu}]_{D}=\delta ^{\mu}~_{\nu}-n^{\mu}n_{\nu},\\
[q^{\mu},q^{\nu}]_{D}=[p^{\mu},p^{\nu}]_{D}=0.
\end{align}
The important result at this point is that in classical mechanics, there is no effect of the \textquoteleft external world\textquoteright \, on the dynamics of our particle \cite{Ogawa}. In the quantum mechanical treatment, the effect of this external world depends on how quantization is performed \cite{Ogawa, Ikegami}. In Dirac's procedure Dirac brackets are replaced by $-i/\hbar $ times commutators, and this is a fixed ingredient of the recipe. However, the ordering of operators is not fixed \cite{Ikegami}, and one needs to impose some ordering. The choice of Ogawa \textit{et. al.} is to write the product of any two operators (notice that this includes simple functions, coordinate transformation matrices, usual momentum operators etc.) in a manifestly symmetric way:
\begin{align}
\lbrace A,B\rbrace =\dfrac{1}{2}(AB+BA).
\end{align}
It is important to stress that this is \textit{hypothesized} in the work of Ogawa \textit{et. al.}; different orderings may also be chosen. 

With this hypothesis and the above information, the Hamiltonian operator takes the following form \cite{Ogawa} (see Appendix B for details):
\begin{align}
H=\dfrac{1}{2}g^{-1/4}p_{i}g^{1/2}g^{ij}p_{j}g^{-1/4}+V(q)+\dfrac{\hbar }{8}g^{00}\Gamma ^{i}_{i0}\Gamma ^{j}_{j0}
\end{align}
with the constraints
\begin{align}
q^{0}=0, \; p_{0}=0.
\end{align}
Here, $g\equiv det(g_{ij})$. This result is obviously different from that of da Costa (and other people who treat the problem in the same way as da Costa). The kinetic term is not simply $-\hbar ^{2}$ times the Laplacian defined on the hypersurface (i.e. $g^{-1/2}p_{i}g^{1/2}g^{ij}p_{j}$); and the additional potential term is also different. The difference between this kinetic term (named as Laplace-Beltrami operator in \cite{Ogawa}) and the usual Laplacian is not a scalar function which may be added to the additional potential, but is another kinetic operator:
\begin{align}
[g^{-1/4},p_{i}g^{1/2}g^{ij}p_{j}g^{-1/4}] & =(-\hbar ^{2} \times \mbox{difference of Laplace-Beltrami and Laplacian)} \nonumber \\
& =\dfrac{i\hbar }{2}g^{-1/4}(g^{1/4}g^{ij}\Gamma ^{k}_{ki}p_{j}+p_{j}g^{1/4}g^{ij}\Gamma ^{k}_{ki}).
\end{align}
Let us also analyze the additional potential terms further. First, notice that there is a difference in the two metric tensors, that of da Costa and Ogawa \textit{et.al}. In order to compare the two additional terms, let us now consider the case in which $g_{00}=1$. Then Ogawa \textit{et.al.}'s additional term becomes $\Gamma ^{i}_{i0}\Gamma ^{j}_{j0}$; but the additional potential derived by da Costa becomes something different. Notice also that, one should consider the connection coefficients in the $q^{3}=0$ limit of da Costa's treatment. Using the metric introduced by da Costa and taking the relevant limit, one obtains (see \ref{connections}):
\begin{align}
& \Gamma ^{k}_{i0}=\alpha ^{k}~_{i}, \\
& \Gamma ^{k}_{k0}=\alpha ^{k}~_{k}=\alpha _{k}~^{k}=Tr(\alpha ),\\
& V_{S}(da \; Costa)\sim 2\Gamma ^{k}_{i0}\Gamma ^{i}_{k0}-\Gamma ^{k}_{k0}\Gamma ^{i}_{i0},
\end{align}
where $\alpha $ is the Weingarten matrix. Notice that $\alpha $ has previously been defined as an object in $E^{3}$. However, the same definition of $\alpha $ applies to the general case, in which the constraint surface is an $N-1$ dimensional hypersurface, having unit normal $\mathbf{n}$, of $E^{N}$. In the above lines we have made use of this fact. The last line is valid only when one studies the problem in $E^{3}$; the reason for this situation is given in the following section. The difference between the two approaches is discussed in the literature (see for example \cite{Ikegami, Ogawatek}), but it is not clear which method is the correct one in a realistic problem.

\subsection{Treatment of Ikegami, Nagaoka, Takagi and Tanzawa}

In this section, we will briefly discuss the idea of Ikegami \textit{et.al.} explained in \cite{Ikegami}. 

The idea proposed by the authors is a simple modification of Ogawa \textit{et.al.}'s treatment. They replace the geometrical constraint $f(x)=0$ (defining an $N-1$ dimnsional hypersurface of $E^{N}$) which is included in the Lagrangian in the previous treatment, with its time derivative set equal to zero:
\begin{align}
& \dot{f}(x)=\dot{x}^{a}\partial _{a}f(x),\\
& L=\dfrac{1}{2}\dot{x}^{a}\dot{x}_{a}-V(x)+\lambda \dot{x}^{a}\partial _{a}f. \label{ikelag}
\end{align}
With this Lagrangian, one obtains the following Hamiltonian and brackets (see Appendix B) \cite{Ikegami}:
\begin{align}
& H=\dfrac{1}{2}p_{a}(\delta ^{ab}-n^{a}n^{b})p_{b}+V(x),\\
& [x^{a},x^{b}]_{D}=[p^{a},p^{b}]_{D}=0,\nonumber \\
& [x^{a},p_{b}]_{D}=\delta ^{a}~_{b}.
\end{align}
Notice that the kinetic part is the square of the momentum tangent to the hypersurface. When the usual replacement of momenta with $-i\hbar $ times partial derivatives is performed, the kinetic part becomes $-\hbar ^{2}$ times the Laplacian of the hypersurface \cite{Golovnev} (this is what Golovnev directly suggests):
\begin{align}
\hat{\nabla }^{2}=(\nabla - \hat{n}(\hat{n}\cdot \nabla))^{2}.
\end{align} 
At this point, the authors make a further modification, and relax the coordinate transformation hypothesis of Ogawa \textit{et.al.}; they suggest considering the most general linear combination of operators at the quantization step \cite{Ikegami}. With this suggestion, one writes the following kinetic term (the authors perform the calculation in a somewhat different way, but our results here will coincide with theirs; see Appendix B for details):
\begin{align}
H=\dfrac{1}{2}p^{a}p_{a}-\dfrac{1}{2}\Bigg(& An^{a}p_{a}n^{b}p_{b}+Bp_{a}n^{a}p_{b}n^{b}+Cp_{a}n^{a}n^{b}p_{b}\nonumber \\
+ & Dn^{a}p_{a}p_{b}n^{b}+Ep_{a}n^{b}p_{b}n^{a}+Fn^{a}p_{b}n^{b}p_{a}\nonumber \\
+ & Gp_{a}p_{b}n^{a}n^{b}+Hn^{a}n^{b}p_{a}p_{b}\Bigg),
\label{Tn}
\end{align}
where the coefficients are arbitrary complex numbers, and the term including these coefficients is defined as $T_{n}$ within the text. Since this operator is part of the Hamiltonian, that is, an observable, it should be Hermitian. This condition gives:
\begin{align}
A^{*}=B, \; C^{*}=C, \; D^{*}=D, \; E^{*}=F, \; G^{*}=H.
\end{align}
Now, in order to make a comparison with Ferrari and Cuoghi's \textquoteleft geometric potential\textquoteright , let us restrict the discussion to three dimensions (i.e. to $E^{3}$) and write the additional potential of da Costa in the following way \cite{Ikegami, Ferrari}:
\begin{align}
V_{s}=-\dfrac{\hbar ^{2}}{2}\left (-\dfrac{1}{4}[Tr(\alpha )]^{2}+\dfrac{1}{2}Tr(\alpha ^{2}) \right )
\label{Costapot}
\end{align}
using the following fact (since $\alpha $ is a $2\times 2$ matrix)
\begin{align}
& det \; \alpha = \dfrac{1}{2!}\epsilon ^{ij}\epsilon _{i'j'}\alpha _{i}~^{i'}\alpha _{j}~^{j'},\\
& \epsilon ^{ij}\epsilon _{i'j'}=\delta ^{i}~_{i'}\delta ^{j}~_{j'}-\delta ^{i}~_{j'}\delta ^{j}~_{i'},\\
& (Tr \; \alpha )^{2}-Tr(\alpha ^{2})=2det\; \alpha .
\label{trdet}
\end{align}
In order to be able to calculate the desired additional potential, one tries to find the necessary conditions on the coefficients. If one also tries to obtain the Laplacian of the hypersurface in the kinetic part, one should try to write (\ref{Tn}) as $\dfrac{1}{2}(p^{a}p_{a}-p_{a}n^{a}n^{b}p_{b})$ plus remaining terms, and try to generate da Costa's potential from the remaining terms. Our results concerning this task has put forward the following conditions on the coefficients (see Appendix B):
\begin{align}
2Re(A)+2Re(G)+2Re(E)+C+D=1,\nonumber \\
A+G^{*}=-D, \; Im(E+G)=0.
\label{hermite}
\end{align}
In our work, we found that Ogawa \textit{et.al.}'s ordering hypothesis does not give the desired results (see Appendix B) and this is in agreement with Ikegami \textit{et. al.}'s result. The authors also assert that there should be some condition imposed on the operators to fix the ordering. However, they take $T_{n}$ as a purely quantum mechanical object, which puts the resulting Hamiltonian in accordance with that of da Costa's \cite{Ikegami}. 

Here, there is another important issue. The comparison of Ogawa \textit{et.al.}'s and da Costa's results leads us to restrict the discussion to $E^{3}$, because the form of the geometrical potential given by da Costa includes $Tr(\alpha )$ and $det(\alpha )$, while the other includes $Tr(\alpha )$ and $Tr(\alpha ^{2} )$, and da Costa's expression is converted into (\ref{Costapot}) by making use of (\ref{trdet}). When higher dimensions are considered, the two expressions may not be comparable in general. For example, in four dimensions (that is, when $\alpha $ is a $3\times 3$ matrix):
\begin{align}
det \; \alpha = \dfrac{1}{3!}\Big((Tr \; \alpha )^{3}-3Tr (\alpha )^{2}Tr \; \alpha +2Tr(\alpha ^{3})\Big),
\label{3dalpha}
\end{align}
making use of
\begin{align}
det \; \alpha = & \dfrac{1}{3!}\epsilon ^{ijk}\epsilon _{i'j'k'}\alpha _{i}~^{i'}\alpha _{j}~^{j'}\alpha _{k}~^{k'},\\
\epsilon ^{ijk}\epsilon _{i'j'k'}= & \delta ^{i}~_{i'}\delta ^{j}~_{j'}\delta ^{k}~_{k'}-\delta ^{i}~_{j'}\delta ^{j}~_{i'}\delta ^{k}~_{k'}+\delta ^{i}~_{j'}\delta ^{j}~_{k'}\delta ^{k}~_{i'}\nonumber \\ - & \delta ^{i}~_{k'}\delta ^{j}~_{j'}\delta ^{k}~_{i'}+\delta ^{i}~_{k'}\delta ^{j}~_{i'}\delta ^{k}~_{j'}-\delta ^{i}~_{i'}\delta ^{j}~_{k'}\delta ^{k}~_{j'}.
\end{align}
These together reveal that the values of the complex coefficients used in (\ref{Tn}) depend on the number of dimensions.   

\section{DIRAC EQUATION}
\label{sec:Dirac Equation}

\subsection{Thin Layer Method Applied to Dirac Equation}
Let us now try to apply the thin layer method to the Dirac equation. Extremizing the following action:
\begin{align}
S=\int d^{4}x\overline{\psi}(i\gamma ^{a}\partial _{a}-m)\psi ,
\end{align}
one obtains Dirac equation in flat spacetime:
\begin{align}
(i\gamma ^{a}\partial _{a}-m)\psi =0,
\end{align}
in natural units and Cartesian coordinates. $\gamma ^{a}$ are vectors under general coordinate transformations and $4 \times 4$ matrices satisfying \cite{Chapman}:
\begin{align}
[\gamma ^{a},\gamma ^{b}]_{+}\equiv \gamma ^{a}\gamma ^{b}+\gamma ^{b}\gamma ^{a}=2\eta ^{ab},
\label{flatalgebra}
\end{align}
where $\eta ^{ab}=diag (1,-1,-1,-1)$. The solution of this equation, $\psi $ is called a four spinor and is a scalar under general coordinate transformations \cite{Bertlmann}.

When one considers an arbitrary curved spacetime, or an arbitrary orthonormal curvilinear coordinate system in flat spacetime, the partial derivative is replaced by some covariant derivative $D_{\mu}$, and the action becomes \cite{Bertlmann}:
\begin{align}
S=\int \overline{\psi}(i\gamma ^{a}E_{a}~^{\mu}D_{\mu}-m)\psi \sqrt{-G}d^{4}q,
\label{curvedaction}
\end{align}
where the quantity $E_{a}~^{\mu}$ will be defined below. 

Let us now consider part of the discussion given in \cite{Chapman}. Under a general coordinate transformation to some set of orthonormal curvilinear frame:
\begin{align}
[\gamma ^{\mu},\gamma ^{\nu}]_{+}=[\dfrac{\partial q^{\mu}}{\partial x^{a}}\gamma ^{a},\dfrac{\partial q^{\nu}}{\partial x^{b}}\gamma ^{b}]_{+}=\dfrac{\partial q^{\mu}}{\partial x^{a}}\dfrac{\partial q^{\nu}}{\partial x^{b}}[\gamma ^{a},\gamma ^{b}]_{+}=2G^{\mu \nu}
\end{align}
is the relevant relation. Notice that since $\gamma ^{a}$ are constant matrices they commute with coordinate transformation matrix elements. One observes here that $\gamma ^{\mu}$ are coordinate dependent. Then, in some orthonormal frame which is not necessarily Cartesian, (\ref{flatalgebra}) is the relevant relation again. But then, in this general curvilinear orthonormal frame, one has:
\begin{align}
e^{a}=e^{a}~_{\mu}dq^{\mu},
\end{align}
satisfying
\begin{align}
& ds^{2}=G_{\mu \nu}dq^{\mu}dq^{\nu}=e^{a}e^{b}\eta _{ab},\\
& e_{a\mu}=\eta _{ab}e^{b}~_{\mu}, \; e^{a\mu}=G^{\mu \nu}e^{a}~_{\nu},
\end{align}
as the basis one forms, which are not in general integrable, that is, they are not exact differentials of any scalar function. Then, one has to introduce a proper connection for the covariant derivative. 

There is one more condition to be met when working with the Dirac equation. One can apply a similarity transformation to the $\gamma $ matrices, and require the equation to remain invariant under such a transformation. This transformation is called a spin transformation, and an equation remaining invariant under such a transformation is called a representation invariant equation \cite{Chapman}. Now, let us consider the flat spacetime Dirac equation in general curvilinear coordinates:
\begin{align} 
(i\gamma ^{\mu}\partial _{\mu}-m)\psi(q)=0.
\end{align}
The spinor is a coordinate scalar, as mentioned above, then the covariant derivative simply becomes a partial derivative. Now applying a similarity transformation to the $\gamma $ matrices:
\begin{align}
& \gamma '^{\mu}\equiv N^{-1}\gamma ^{\mu}N,\\
\Rightarrow & \psi \rightarrow \psi '\equiv N^{-1}\psi ,\\
\Rightarrow & (iN^{-1}\gamma ^{\mu}N\partial _{\mu}-m)N^{-1}\psi =N^{-1}\Bigg(i\gamma ^{\mu}\Big(\partial _{\mu}+NN^{-1}_{,\mu}\Big)-m\Bigg)\psi =0.
\end{align}
Here $N$ is also a coordinate scalar. One immediately observes that the form of the equation is now different, that is, the equation has not remained invariant. In order to overcome this problem, one introduces the so called spin covariant derivative with its proper connection which leaves the equation invariant under both general coordinate and spin transformations:
\begin{align}
D_{\mu}\psi (q)=\psi (q)_{;\mu}+\Gamma _{\mu}(q)\psi (q),
\end{align}
where $\Gamma _{\mu}$ are called Fock-Ivanenko coefficients. Although this representation will not be used in the following parts, the above discussion followed from \cite{Chapman} briefly gives the reasoning which lies under the requirements that one should impose on the equation. The proper covariant derivative which is used in  our work has the following connection \cite{Bertlmann}:
\begin{align}
& \omega _{\mu}=\dfrac{1}{8}\omega _{ab\mu}[\gamma ^{a},\gamma ^{b}],\\
& \omega ^{a}~_{b\mu}\equiv -E_{b}~^{\nu}\nabla _{\mu}e^{a}~_{\nu}=-E_{b}~^{\nu}(\partial _{\mu}e^{a}~_{\nu}-\Gamma^{\lambda}_{\mu \nu}e^{a}~_{\lambda}),
\end{align}
where
\begin{align}
& ds^{2}=e^{a}~_{\mu}e^{b}~_{\nu}dq^{\mu}dq^{\nu},\\
& e^{a}~_{\mu}E_{b}~^{\mu}=\delta ^{a}~_{b}, \; e^{a}~_{\mu}E_{a}~^{\nu}=\delta ^{\nu}~_{\mu}.
\end{align}
$e^{a}~_{\mu}$ are called vierbeins (vielbein in general dimensions).

Now, let us try to apply the thin layer method to the Dirac equation. We assume that (similar to \cite{Ferrari, daCosta}), our particle is constrained to move on a surface $f(x,y,z)=0$. The metric in curvilinear coordinates is then:
\begin{eqnarray}
[G_{\mu \nu}]=\begin{bmatrix}
1 & 0 & 0\\
0 & -G_{ij} & 0\\
0 & 0 & -1 \end{bmatrix}
\end{eqnarray}
where the spatial part of the metric is $-1$ times the metric introduced by Ferrari and Cuoghi up to $O(q^{3})$. The reason is that Dirac equation has only first order derivatives, and an expansion to first order in $q^{3}$ will be enough. Then, we introduce the vierbeins:
\begin{align}
e^{a}~_{i}=\partial_{i} x^{a}+q^{3}H_{ij}g^{jk}\partial _{k}x^{a}, \;  e^{0}~_{0}=1, \; e^{a}~_{3}=N^{a}, \; \mbox{others}=0
\end{align}
where the subscript zero denotes time coordinate and $N^{a}$ are the Cartesian components of the surface normal (which is a unit vector), just as in Chapter 2. These give the inverse vierbeins:
\begin{align}
E_{a}~^{i}=\eta _{ab}(g^{ij}-q^{3}H^{ij})\partial _{j}x^{b}, \; E_{0}~^{0}=1, \; E_{a}~^{3}=N_{a}, \; \mbox{others}=0. 
\end{align}
In order to calculate the spin connection coefficients, Christoffel symbols are to be calculated. They have already been in service, calculated in Chapter 2, equations (\ref{connections}). Using these and the definition of spin connection, one obtains:
\begin{align} 
& \omega ^{c}~_{d0}=0,\nonumber \\
& \omega ^{c}~_{d3}=0,\nonumber \\
& \omega ^{c}~_{di}=\tilde{\omega} ^{c}~_{di}-N^{c}H_{ki}\tilde{E}_{d}~^{k}+O(q^{3}),
\end{align}
by direct calculation. Since the connection coefficients are not differentiated, one does not need the explicit expression of $O(q^{3})$ term. The over tildes indicate that the quantity is calculated using only surface coordinates, that is:
\begin{align}
& \tilde{E}_{d}~^{k}=\eta _{da}g^{kl}\tilde{e}^{a}~_{l},\nonumber \\
& \tilde{\omega} ^{c}~_{di}=-\tilde{E}_{d}~^{k}(\partial _{i}\tilde{e}^{c}~_{k}-\tilde{\Gamma}^{l}_{ki}\tilde{e}^{c}~_{l}),\nonumber \\
& \tilde{\Gamma}^{l}_{ki}=\dfrac{1}{2}g^{lj}(g_{lk,i}+g_{li,k}-g_{ki,l}).
\end{align}
Next, one rescales $\psi $ as in Chapter 2. This has been done in another article \cite{Burgess}, and the rescaled spinor is:
\begin{align}
& \chi \equiv \psi \sqrt{F},\nonumber \\
& F\equiv 1+q^{3}Tr \; \alpha .
\end{align}
Notice that the rescaling factor is $O(q^{3})$ this time. Now, varying the action (\ref{curvedaction}) with respect to $\overline{\psi}$, one obtains the curved spacetime Dirac equation:
\begin{align}
(i\gamma ^{a}E_{a}~^{\mu}D_{\mu}-m)\psi =0.
\label{curvedDirac}
\end{align}
Putting everything into (\ref{curvedDirac}), one obtains:
\begin{align}
& \Bigg(i\gamma ^{0}\partial _{0}\chi +\dfrac{i}{8}\gamma ^{a}\tilde{E}_{a}~^{i}\eta _{bc}(\tilde{\omega}^{c}~_{di}-N^{c}H_{ki}\tilde{E}_{d}~^{k})[\gamma ^{b},\gamma ^{d}]\chi \nonumber \\
& + i\gamma ^{a}\tilde{E}_{a}~^{i}\partial _{i}\chi + i\gamma ^{a}N_{a}(\partial _{3}\chi -\dfrac{1}{2}(Tr\; \alpha ) \chi ) -m\chi \Bigg)_{q^{3}=0}=0
\end{align}
in the $q^{3}=0$ limit. Notice that this equation is also separable, that is, dynamics involving 2D surface coordinates and the coordinate orthogonal to the 2D surface can be decoupled. But notice also that, the representation of Dirac matrices is still 4 dimensional, so one may search for an alternative representation which is indeed written in 3 spacetime dimensions \cite{Burgess}.  

Another observation one makes here is that this equation includes an additional potential term, $-\dfrac{i}{2}\gamma ^{a}N_{a}Tr\; \alpha $ in natural units. This potential arises from the rescaling of the 4 spinor $\psi $. There is one other additional term in the equation, coming from the connection coefficients: $\dfrac{i}{8}\gamma ^{a}\tilde{E}_{a}~^{i}\eta _{bc}N^{c}H_{ki}\tilde{E}_{d}~^{k}[\gamma ^{b},\gamma ^{d}]$. This term can also be interpreted as an additional potential term, but whether this term is a real scalar or not is to be verified. Such a term did not appear in the treatment for Schr\"{o}dinger equation. This difference arises from the difference between the covariant derivatives. 

\section{CONCLUSION}
\label{sec:Conclusion}
In our work, we have analyzed two different approaches \cite{Ferrari, Ogawa}, that of da Costa and Dirac, and two variants of Dirac's approach \cite{Ikegami}. Another approach, that of Golovnev, has not been analyzed but derived from the so called \textquoteleft conservative constraint condition\textquoteright \, of Ikegami (see \cite{Golovnev, Ikegami}). We have also discussed how Dirac's approach can be adjusted to give the same result with da Costa. The reason for doing so is that da Costa's approach is rather geometrical, and is independent from the quantization procedure; that is, whatever the wave equation is, one can expand the equation in powers of one coordinate (which essentially is small compared to the length scales relevant to the system) and calculate the limiting equation as this coordinate approaches to zero. This fact makes this approach to be closer to a correct description; whatever that description is. However, Dirac's procedure is algebraic, and involves the canonical formalism of classical mechanics and canonical quantization. Such a procedure seems to be necessary for dealing with constrained quantum systems, since the experimentally verified applications of quantum theory are based on certain quantization procedures. Dirac's work reveals some ambiguities in such a trial \cite{Dirac, Ikegami}. As stated above, the geometrical approach of da Costa seems to be closer to give the correct result, due to being a direct expansion of the relevant Schr\"{o}dinger equation; but the existence of geometrical constraints inevitably imply physical interactions, for it is impossible to constrain a system geometrically without making use of interactions; and as a result, Dirac's procedure may predict different aspects of such an interaction which may have not been taken into account before. 

We have also tried to calculate the spin-magnetic field interaction and connection coefficients of the Dirac equation using da Costa's approach, which we have not been able to find in the literature. But this discussion requires a more detailed analysis, which will possibly be given in the future. One should also apply Dirac's method to write a constrained Dirac equation, to compare the results with those of da Costa's approach, and possible interactions which can effectively be viewed as constraining mechanisms should be considered.

\appendix
\section{APPENDIX}

\subsection{Gauge Invariance of (\ref{schrodinger})}
In \cite{Ferrari}, the following well known gauge transformation is mentioned:
\begin{align}
A'_{\mu}\equiv A_{\mu}+\partial _{\mu}\gamma , \; A'_{0}\equiv A_{0}+\partial _{0}\gamma , \; \psi '\equiv \psi \exp (\dfrac{iQ\gamma}{\hbar}),
\end{align}
where the subscript zero denotes time derivative and $\gamma (t,q^{\mu})$ is some scalar function. Putting these new objects directly into (\ref{schrodinger}) and taking the derivatives, one obtains the following:
\begin{align}
& i\hbar \partial _{0}\Big(\Psi \exp (\dfrac{iQ\gamma}{\hbar})\Big)+QA_{0}\Psi \exp (\dfrac{iQ\gamma}{\hbar})+i\hbar \partial _{0}\gamma  \Psi \exp (\dfrac{iQ\gamma}{\hbar})=\nonumber \\
& \dfrac{1}{2m}\Bigg[-\dfrac{\hbar ^{2}}{\sqrt{G}}\partial _{\mu}\Big(\sqrt{G}G^{\mu \nu}\partial _{\nu}\Big(\Psi \exp (\dfrac{iQ\gamma}{\hbar})\Big)\Big) +
 \dfrac{iQ\hbar}{\sqrt{G}}\partial _{\mu}\Big(\sqrt{G}G^{\mu \nu}\Big(A_{\nu}+ \partial _{\nu}\gamma \Big)\Psi \exp (\dfrac{iQ\gamma}{\hbar})\Big)\nonumber \\ 
 & + 2iQ\hbar G^{\mu \nu}\Big(A_{\nu}+\partial _{\nu}\gamma \Big)\partial _{\mu}\Big(\Psi \exp (\dfrac{iQ\gamma}{\hbar})\Big) +
 Q^{2}G^{\mu \nu}\Big(A_{\mu}+\partial _{\mu}\gamma \Big)\Big(A_{\nu}+\partial _{\nu}\gamma \Big)\Psi \exp (\dfrac{iQ\gamma}{\hbar})\Bigg],
\end{align}
where $A_{0}=-V$. Now, expanding derivatives, one obtains:
\begin{align}
& i\hbar \partial _{0}\Big(\Psi \exp (\dfrac{iQ\gamma}{\hbar})\Big)=i\hbar (\partial _{0}\Psi )\exp (\dfrac{iQ\gamma}{\hbar})-i\hbar \psi Q(\partial _{0}\gamma ) \exp(\dfrac{iQ\gamma}{\hbar}),\\
& \partial _{\mu}\Big(\psi \exp (\dfrac{iQ\gamma}{\hbar})\Big)=\exp (\dfrac{iQ\gamma}{\hbar})\partial _{\mu}\psi +\dfrac{iQ}{\hbar}\partial _{\mu}\gamma (\dfrac{iQ\gamma}{\hbar})\psi ,\\
& \partial _{\mu}\Big(\sqrt{G}G^{\mu \nu}\partial _{\nu}\Big(\Psi \exp (\dfrac{iQ\gamma}{\hbar})\Big)\Big)=\Bigg(\partial _{\mu}\Big(\sqrt{G}G^{\mu \nu}\partial _{\nu}\Psi \Big) +2 \dfrac{iQ}{\hbar}\partial _{\mu}\Big(\sqrt{G}G^{\mu \nu}\Psi \Big)\partial _{\nu}\gamma \nonumber \\
& + \dfrac{iQ}{\hbar}\sqrt{G}G^{\mu \nu}\Psi \partial _{\mu}\partial _{\nu}\gamma  - \dfrac{Q^{2}}{\hbar ^{2}}\sqrt{G}G^{\mu \nu}\Psi \partial _{\mu}\gamma \partial _{\nu}\gamma \Bigg)\exp (\dfrac{iQ\gamma}{\hbar}),\\
& \partial _{\mu}\Big(\sqrt{G}G^{\mu \nu}\Big(A_{\nu}+ \partial _{\nu}\gamma \Big)\Psi \exp (\dfrac{iQ\gamma}{\hbar})\Big)=\exp (\dfrac{iQ\gamma}{\hbar})
\Bigg(\partial _{\mu}\Big(\sqrt{G}G^{\mu \nu}A_{\nu}\psi \Big)+\dfrac{iQ}{\hbar}\sqrt{G}G^{\mu \nu}A_{\nu}\psi \partial _{\mu}\gamma \nonumber \\
& + \partial _{\mu}\Big(\sqrt{G}G^{\mu \nu}\psi \Big)\partial _{\nu}\gamma  + \sqrt{G}G^{\mu \nu}\psi \partial _{\mu}\partial _{\nu}\gamma + \dfrac{iQ}{\hbar} \sqrt{G}G^{\mu \nu}\psi \partial _{\mu}\gamma \partial _{\nu}\gamma \Bigg).
\end{align}
Then, putting the above terms into (\ref{schrodinger}) and cancelling exponentials from both sides, one directly obtains (\ref{schrodinger}) again.

\subsection{Dirac Brackets and (\ref{Pois1})-(\ref{Pois4})}
We already know that Poisson brackets satisfy those four important properties. Then, all Poisson bracket terms in the definition of Dirac brackets satisfy (\ref{Pois1}). The matrix entries $\Delta _{kl}=[\phi _{k},\phi _{l}]$ also satisfy (\ref{Pois1}). Then, both terms present in the definition of Dirac brackets satisfy this, so Dirac brackets satisfy (\ref{Pois1}). 

For (\ref{Pois2}), let us write the expression explicitly:
\begin{align}
[f+g,h]^{*} & =[f+g,h]-[f+g,\phi _{k}]\Delta ^{-1}_{kl}[\phi _{l},h]\nonumber \\
& =[f,h]-[f,\phi _{k}]\Delta ^{-1}_{kl}[\phi _{l},h]+[g,h]-[g,\phi _{k}]\Delta ^{-1}_{kl}[\phi _{l},h]=[f,h]^{*}+[g,h]^{*}.
\end{align}

For (\ref{Pois3}), let us again write the expression explicitly:
\begin{align}
[fg,h]^{*} & =[fg,h]-[fg,\phi _{k}]\Delta ^{-1}_{kl}[\phi _{l},h]\nonumber \\
& =[f,h]g+f[g,h]-\left([f,\phi _{k}]g+f[g,\phi _{k}]\right)\Delta ^{-1}_{kl}[\phi _{l},h]=[f,h]^{*}g+f[g,h]^{*}.
\end{align}

For (\ref{Pois4}), it is useful to follow the direct proof given in the Appendix of \cite{Diracham}:
\begin{align}
\left[f,[g,h]^{*}\right]^{*} & =\left[f,[g,h]^{*}\right]-[f,\phi _{m}]\Delta ^{-1}_{mn}\left[\phi _{n},[g,h]^{*}\right]\nonumber \\
& =\left[f,[g,h]\right]-\left[f,[g,\phi _{k}]\Delta ^{-1}_{kl}[\phi _{l},h]\right]-[f,\phi _{m}]\Delta ^{-1}_{mn}\Bigg(\Big[\phi _{n},[g,h]\Big]-\Big[\phi _{n},[g,\phi _{k}]\Delta ^{-1}_{kl}[\phi _{l},h]\Big]\Bigg)\nonumber \\
& =[f,[g,h]]-[f,[g,\phi _{k}]]\Delta ^{-1}_{kl}[\phi _{l},h]-[f,\Delta ^{-1}_{kl}][g,\phi _{k}][\phi _{l},h]\nonumber \\
& -[f,[\phi _{l},h]]\Delta ^{-1}_{kl}[g,\phi _{k}]-[f,\phi _{m}]\Delta ^{-1}_{mn}[\phi _{n},[g,h]]\nonumber \\
& +[f,\phi _{m}]\Delta ^{-1}_{mn}[\phi _{n},[g,\phi _{k}]]\Delta ^{-1}_{kl}[\phi _{l},h]+[f,\phi _{m}]\Delta ^{-1}_{mn}[\phi _{n},\Delta ^{-1}_{kl}][g,\phi _{k}][\phi _{l},h]\nonumber \\
& +[f,\phi _{m}]\Delta ^{-1}_{mn}[\phi _{n},[\phi _{l},h]]\Delta ^{-1}_{kl}[g,\phi _{k}].
\label{ilkifade}
\end{align}
It is obvious that the first term, which is an ordinary Poisson bracket, satisfies the Jacobi identity. Now, let us consider the second, fourth and fifth terms. When $f,g$ and $h$ are permuted cyclically, and the nine terms are summed, one obtains:
\begin{align}
& [h,\phi _{k}]\Delta ^{-1}_{kl}\Bigg([f,[g,\phi _{l}]]+[\phi _{l},[f,g]]+[g,[\phi _{l},f]]\Bigg)+[g,\phi _{k}]\Delta ^{-1}_{kl}\Bigg([h,[f,\phi _{l}]]+[\phi _{l},[h,f]]+[f,[\phi _{l},h]]\Bigg)\nonumber \\
+ & [f,\phi _{k}]\Delta ^{-1}_{kl}\Bigg([g,[h,\phi _{l}]]+[\phi _{l},[g,h]]+[h,[\phi _{l},g]]\Bigg)=0.
\end{align}
Now, consider the sixth and eighth terms (with $f,g,h$ permuted cyclically and the six terms summed):
\begin{align}
& [f,\phi _{m}]\Delta ^{-1}_{mn}\Delta ^{-1}_{kl}[h,\phi _{l}]\Bigg([\phi _{n},[g,\phi _{k}]]+[\phi _{k},[\phi _{n},g]]\Bigg)\nonumber \\
+ & [g,\phi _{m}]\Delta ^{-1}_{mn}\Delta ^{-1}_{kl}[f,\phi _{l}]\Bigg([\phi _{n},[h,\phi _{k}]]+[\phi _{k},[\phi _{n},h]]\Bigg)\nonumber \\
+ & [g,\phi _{m}]\Delta ^{-1}_{mn}\Delta ^{-1}_{kl}[g,\phi _{l}]\Bigg([\phi _{n},[f,\phi _{k}]]+[\phi _{k},[\phi _{n},f]]\Bigg)\nonumber \\
= & -[f,\phi _{m}]\Delta ^{-1}_{mn}\Delta ^{-1}_{kl}[h,\phi _{l}][g,[\phi _{k},\phi _{n}]] - [g,\phi _{m}]\Delta ^{-1}_{mn}\Delta ^{-1}_{kl}[f,\phi _{l}][h,[\phi _{k},\phi _{n}]] - \nonumber \\
& [h,\phi _{m}]\Delta ^{-1}_{mn}\Delta ^{-1}_{kl}[g,\phi _{l}][f,[\phi _{k},\phi _{n}]],
\end{align}
using the Jacobi identity. Remembering that
\begin{align}
& \Delta ^{-1}_{kl}[\phi _{n},\phi _{k}]=\delta _{nl},\nonumber \\
& [F(q,p),\delta _{nl}]=0,\nonumber \\
\Rightarrow & [F,\Delta ^{-1}_{kl}][\phi _{n},\phi _{k}]=-[F,[\phi _{n},\phi _{k}]]\Delta ^{-1}_{kl},
\end{align}
where one uses the Jacobi identity while passing to the last line. Then one obtains:
\begin{align}
& -[f,\phi _{m}]\Delta ^{-1}_{mn}\Delta ^{-1}_{kl}[h,\phi _{l}][g,[\phi _{k},\phi _{n}]] - [g,\phi _{m}]\Delta ^{-1}_{mn}\Delta ^{-1}_{kl}[f,\phi _{l}][h,[\phi _{k},\phi _{n}]] - \nonumber \\
& [h,\phi _{m}]\Delta ^{-1}_{mn}\Delta ^{-1}_{kl}[g,\phi _{l}][f,[\phi _{k},\phi _{n}]] = -[f,\phi _{k}][h,\phi _{l}][g,\Delta ^{-1}_{kl}]-[g,\phi _{k}][f,\phi _{l}][h,\Delta ^{-1}_{kl}]\nonumber \\
& -[h,\phi _{k}][g,\phi _{l}][f,\Delta ^{-1}_{kl}].
\end{align}
Notice that when $f,g,h$ are permuted cyclically in the third term of (\ref{ilkifade}) and summed, the resulting expression cancels the above expression. 

We are left with the seventh term of (\ref{ilkifade}). Permuting $f,g,h$ cyclically and summing the three terms, one obtains:
\begin{align}
& [f,\phi _{m}]\Delta ^{-1}_{mn}[\phi _{n},\Delta ^{-1}_{kl}][g,\phi _{k}][\phi _{l},h]+[g,\phi _{m}]\Delta ^{-1}_{mn}[\phi _{n},\Delta ^{-1}_{kl}][h,\phi _{k}][\phi _{l},f]\nonumber \\
+ & [h,\phi _{m}]\Delta ^{-1}_{mn}[\phi _{n},\Delta ^{-1}_{kl}][f,\phi _{k}][\phi _{l},g]\nonumber \\
= & [f,\phi _{m}][g,\phi _{k}][\phi _{l},h]\Bigg(\Delta ^{-1}_{mn}[\phi _{n},\Delta ^{-1}_{kl}]+\Delta ^{-1}_{ln}[\phi _{n},\Delta ^{-1}_{mk}]+\Delta ^{-1}_{kn}[\phi _{n},\Delta ^{-1}_{lm}]\Bigg).
\end{align}
Now, remembering that $[F(q,p),\Delta ^{-1}_{mk}][\phi _{k},\phi _{a}]=-[F(q,p),[\phi _{k},\phi _{a}]]\Delta ^{-1}_{mk}$, one gets:
\begin{align}
\Delta ^{-1}_{ab}[F(q,p),\Delta ^{-1}_{mk}][\phi _{k},\phi _{a}]=[F(q,p),\Delta ^{-1}_{mb}]=-[F(q,p),[\phi _{k},\phi _{a}]]\Delta ^{-1}_{mk}\Delta ^{-1}_{ab}.
\end{align} 
Then, putting $F(q,p)=\phi _{n}$:
\begin{align}
& \Delta ^{-1}_{mn}[\phi _{n},\Delta ^{-1}_{kl}]+\Delta ^{-1}_{ln}[\phi _{n},\Delta ^{-1}_{mk}]+\Delta ^{-1}_{kn}[\phi _{n},\Delta ^{-1}_{lm}]\nonumber \\
= & -[\phi _{n},[\phi _{a},\phi _{b}]]\Delta ^{-1}_{ka}\Delta ^{-1}_{bl}\Delta ^{-1}_{mn}-[\phi _{n},[\phi _{a},\phi _{b}]]\Delta ^{-1}_{ma}\Delta ^{-1}_{bk}\Delta ^{-1}_{ln}-[\phi _{n},[\phi _{a},\phi _{b}]]\Delta ^{-1}_{la}\Delta ^{-1}_{bm}\Delta ^{-1}_{kn}.
\end{align}
Since Jacobi identity applies to Poisson brackets, one can write the equivalent of $[\phi _{n},[\phi _{a},\phi _{b}]]$ using the identity. But for $[\phi _{n},[\phi _{a},\phi _{b}]]$, applying the Jacobi identity is equivalent to permuting $n,a,b$ in a cyclic order. So, using the Jacobi identity for $[\phi _{n},[\phi _{a},\phi _{b}]]$ in one of the terms on the left hand side in this manner, one sees that the expression equals to zero. This completes the proof of (\ref{Pois4}). 

\subsection{Dirac Procedure with Geometrical Constraint}
In order to make calculations one needs to calculate the Dirac brackets first. Using the usual transition to Hamiltonian dynamics, one obtains:
\begin{align}
H=\dfrac{1}{2m}p^{a}p_{a}+V(x)-\lambda f(x).
\end{align}
The procedure requires adding the primary constraints to this Hamiltonian and defining a new Hamiltonian $H^{*}$:
\begin{align}
H^{*}=H+up_{\lambda },
\end{align}
where $u$ is some multiplier. Then, one needs to find the consistency conditions by calculating Poisson brackets of the primary constraints with this $H^{*}$ \cite{Dirac, Ogawa}:
\begin{align}
& \phi _{1}\equiv p_{\lambda }\approx 0,\\
& \phi _{2}\equiv [\phi _{1},H^{*}]=[p_{\lambda },H^{*}]=-\dfrac{\partial H^{*}}{\partial \lambda}=f(x)\approx 0,\\
& \phi _{3}\equiv [\phi _{2},H^{*}]=[f(x),H^{*}]=\partial _{a}f(x)\dfrac{\partial H^{*}}{\partial p_{a}} =p^{a}\partial _{a}f(x)\approx 0,\\
& \phi _{4}\equiv [\phi _{3},H^{*}]=[p^{a}\partial _{a}f(x),H^{*}]=p^{a}p^{b}\partial _{a}\partial _{b}f(x)-\partial _{b}\Big(V(x)-\lambda f(x)\Big)\partial ^{b}f(x)\approx 0.
\end{align}
The procedure terminates here since the last constraint includes $\lambda $ and calculating the Poisson bracket of this with $H^{*}$ and (weakly) equating to zero will bring a condition on $u$. Then, there are four constraints in total. Now, Poisson brackets of the constraints with each other are to be calculated. Nonzero brackets are:
\begin{align}
& [\phi _{1},\phi _{4}]=-\partial _{b}\partial ^{b}f(x),\\
& [\phi _{2},\phi _{3}]=\partial _{b}\partial ^{b}f(x),\\
& [\phi _{2},\phi _{4}]=p^{b}\partial _{b}\Big(\partial _{c}f(x)\partial ^{c}f(x)\Big),\\
& [\phi _{3},\phi _{4}]=2p^{a}p^{c}\Big(\partial _{b}\partial _{a}f(x)\Big)\Big(\partial ^{b}\partial _{c}f(x)\Big)-\partial ^{a}f(x)\partial _{a}\Bigg(p^{a}p^{b}\partial _{a}\partial _{b}f(x)\nonumber \\
&\qquad \quad - \partial _{b}\Big(V(x)-\lambda f(x)\Big)\partial ^{b}f(x)\Bigg).
\end{align}
Like Ogawa \textit{et. al.}, let us define:
\begin{align}
[\phi _{2},\phi _{3}]\equiv \alpha , [\phi _{2},\phi _{4}]\equiv -\beta , [\phi _{4},\phi _{3}]\equiv \gamma . 
\end{align}
Then, one constructs the following matrix and calculates its inverse to find the Dirac brackets \cite{Dirac, Ogawa}:
\begin{eqnarray}
[\Delta _{kl}] =\begin{bmatrix}
0 & 0 & 0 & -\alpha \\
0 & 0 & \alpha & -\beta \\
0 & -\alpha & 0 & -\gamma \\
\alpha & \beta & \gamma & 0 \end{bmatrix}\; ,
\end{eqnarray}
with the inverse:
\begin{eqnarray} 
[\Delta ^{-1}_{kl}]=\begin{bmatrix} 
0 & -\dfrac{\gamma }{\alpha ^{2}} & \dfrac{\beta }{\alpha ^{2}} & \dfrac{1}{\alpha } \\
\dfrac{\gamma }{\alpha ^{2}} & 0 & -\dfrac{1}{\alpha } & 0 \\
-\dfrac{\beta }{\alpha ^{2}} & \dfrac{1}{\alpha } & 0 & 0 \\
-\dfrac{1}{\alpha } & 0 & 0 & 0
\end{bmatrix}\; .
\end{eqnarray}
Now, employing the definiton (\ref{dbdef}) of Dirac brackets, one obtains the following fundamental bracket relations \cite{Ogawa}:
\begin{align}
& [x^{a},p_{b}]_{D}=\delta ^{a}~_{b}-n^{a}n_{b},\\
& [x^{a},x^{b}]_{D}=0,\\
& [p_{a},p_{b}]_{D}=p^{c}(n_{b}\partial _{c}n_{a}-n_{a}\partial _{c}n_{b}).
\end{align}
Notice that Ogawa \textit{et.al.}'s notation for Dirac brackets is used again. The classical (Lagrangian) equations of motion are:
\begin{align}
& \ddot{x}_{a}=-\partial _{a}V(x)+\lambda \partial _{a}f(x),\\
& f(x)=0.
\end{align}
Notice that the second equation is obtained from $\dfrac {\partial L}{\partial \lambda }-\dfrac{d}{dt}(\dfrac {\partial L}{\partial \dot{\lambda }})=0$. Since Dirac brackets are defined, all constraint equations can now be treated as strong equations. Then, the Hamiltonian equations of motion in terms of Dirac brackets read:
\begin{align}
& \dot{x}^{a}=[x^{a},H]_{D}=p^{b}(\delta ^{a}~_{b}-n^{a}n_{b})=p^{a},\\
& \dot{p}_{a}=[p_{a},H]_{D}=p^{b}p^{c}(n_{b}\partial _{c}n_{a}-n_{a}\partial _{c}n_{b})+(n_{a}n^{d}-\delta ^{d}~_{a})\partial _{d}V(x)=(n_{a}n^{d}-\delta ^{d}~_{a})\partial _{d}V(x),\\
& \ddot{x}^{a}=[\dot{x}^{a},H]_{D}=[p^{a},H]_{D}=(n_{a}n^{d}-\delta ^{d}~_{a})\partial _{d}V(x),
\end{align}
using $p_{a}n^{a}=0$. To see the equivalance of those equations, one can solve $\phi _{4}=0$ for $\lambda $ (using $p^{a}\partial _{b}\partial _{a}f=\partial _{b}(p^{a}\partial _{a}f)=0$, since $p^{a}n_{a}=0$ from $\phi _{2}=0$, and the definition of $n^{a}$) and get:
\begin{align}
& \lambda =\dfrac{\partial _{b}V(\partial ^{b}f)}{\partial _{c}f(\partial ^{c}f)},\\
& \ddot{x}_{a}=-\partial _{a}V(x)+\lambda \partial _{a}f(x)=-\partial _{a}V(x)+n_{a}n^{b}\partial _{b}f(x).
\end{align} 
After the coordinate transformation, the brackets become (at the classical level):
\begin{align}
& [q^{\mu},q^{\nu}]_{D}=\dfrac{\partial q^{\mu}}{\partial x^{a}}\dfrac{\partial q^{\nu}}{\partial x^{b}}[x^{a},x^{b}]_{D}=0,\\
& [q^{\mu},p_{\nu}]_{D}=\dfrac{\partial q^{\mu}}{\partial x^{a}}[x^{a},\dfrac{\partial x^{b}}{\partial q^{\nu}}p_{b}]_{D}=\delta ^{\mu}~_{\nu}-n^{\mu}n_{\nu},\\
& [p_{\mu},p_{\nu}]_{D}=[\dfrac{\partial x^{a}}{\partial q^{\mu}}p_{a},\dfrac{\partial x^{b}}{\partial q^{\nu}}p_{b}]_{D}\nonumber \\
& =\dfrac{\partial x^{a}}{\partial q^{\mu}}[p_{a},\dfrac{\partial x^{b}}{\partial q^{\nu}}]_{D}p_{b}+[\dfrac{\partial x^{a}}{\partial q^{\mu}},p_{b}]_{D}p_{a}\dfrac{\partial x^{b}}{\partial q^{\nu}}+\dfrac{\partial x^{b}}{\partial q^{\nu}}\dfrac{\partial x^{a}}{\partial q^{\mu}}[p_{a},p_{b}]=0.
\end{align}
To see that $[p_{\mu},p_{\nu}]_{D}=0$, first notice that:
\begin{align}
& [F(q),p_{\mu}]_{D}=\partial _{\mu}F(q)[q^{\mu},p_{\mu}]-\partial _{\mu}F(q)[q^{\mu},\phi _{k}]\Delta ^{-1}_{kl}[\phi _{l},p_{\nu}]=\partial _{\mu}F(q)[q^{\mu},p_{\nu}]_{D}, \label{turevde}\\
& \dfrac{\partial }{\partial x^{b}}\Big(\dfrac{\partial x^{a}}{\partial q^{\mu}}\Big)
=\dfrac{\partial q^{\lambda }}{\partial x^{b}}\dfrac{\partial }{\partial q^{\lambda }}\Big(\dfrac{\partial x^{a}}{\partial q^{\mu}}\Big).
\label{ozellik}
\end{align}
Using (\ref{ozellik}), the first and second terms cancel each other. In the third term, $[p_{a},p_{b}]_{D}$ is antisymmetric in $a$ and $b$, but the coordinate transformation matrix elements are symmetric, so this term is zero, meaning that the whole expression equals to zero.

At this point quantization is performed; that is, Dirac brackets, satisfying the defining properties and (\ref{turevde}), are replaced by $-i/\hbar $ times commutators. Making use of the coordinate transformation hypothesis of Ogawa \textit{et. al.}, one obtains (using $p_{a}n^{a}\rightarrow \lbrace p_{a},n^{a} \rbrace$):
\begin{align}
& [q^{\mu},q^{\nu}]=0,\\
& [q^{\mu},p_{\nu}]=[q^{\mu},\lbrace \dfrac{\partial x^{a}}{\partial q^{\nu}},p_{a}\rbrace ]=\dfrac{1}{2}\Big(\dfrac{\partial x^{a}}{\partial q^{\nu}}[q^{\mu},p_{a}]+[q^{\mu},p_{a}]\dfrac{\partial x^{a}}{\partial q^{\nu}} \Big ) \nonumber \\
& = \dfrac{\partial x^{a}}{\partial q^{\nu}}\dfrac{\partial q^{\mu}}{\partial x^{b}}[x^{b},p_{a}]=i\hbar (\delta ^{\mu}~_{\nu}-n^{\mu}n_{\nu}) \\
& [p_{\mu},p_{\nu}]=[\lbrace \dfrac{\partial x^{a}}{\partial q^{\mu}},p_{a}\rbrace ,\lbrace \dfrac{\partial x^{b}}{\partial q^{\nu}},p_{b}\rbrace]\nonumber \\
& = \dfrac{1}{4}\Bigg(\dfrac{\partial x^{b}}{\partial q^{\nu}}[\dfrac{\partial x^{a}}{\partial q^{\mu}},p_{b}]p_{a}+\dfrac{\partial x^{a}}{\partial q^{\mu}}\Big([p_{a},\dfrac{\partial x^{b}}{\partial q^{\nu}}]p_{b}+\dfrac{\partial x^{b}}{\partial q^{\nu}}[p_{a},p_{b}]\Big)+\Big([p_{a},\dfrac{\partial x^{b}}{\partial q^{\nu}}]p_{b} \nonumber \\
& +\dfrac{\partial x^{b}}{\partial q^{\nu}}[p_{a},p_{b}]\Big)\dfrac{\partial x^{a}}{\partial q^{\mu}}+p_{a}\dfrac{\partial x^{b}}{\partial q^{\nu}}[\dfrac{\partial x^{a}}{\partial q^{\mu}},p_{b}]+ [\dfrac{\partial x^{a}}{\partial q^{\mu}},p_{b}]\dfrac{\partial x^{b}}{\partial q^{\nu}}p_{a}+\dfrac{\partial x^{a}}{\partial q^{\mu}}\Big([p_{a},p_{b}]\dfrac{\partial x^{b}}{\partial q^{\nu}}\nonumber \\
& +p_{b}[p_{a},\dfrac{\partial x^{b}}{\partial q^{\nu}}]\Big)+\Big([p_{a},p_{b}]\dfrac{\partial x^{b}}{\partial q^{\nu}}+p_{b}[p_{a},\dfrac{\partial x^{b}}{\partial q^{\nu}}]\Big)\dfrac{\partial x^{a}}{\partial q^{\mu}}+p_{a}[\dfrac{\partial x^{a}}{\partial q^{\mu}},p_{b}]\dfrac{\partial x^{b}}{\partial q^{\nu}}\Bigg)=0.
\label{pmupnu}
\end{align} 
Within the last equation, there are some obvious cancellations. Using (\ref{ozellik}), the first term cancels the second, and the eleventh term cancels the twelfth. Since $[p_{a},p_{b}]$ is antisymmetric in $a$ and $b$, but the coordinate transformation matrix elements are symmetric, third and tenth terms vanish. Then, one realizes that:
\begin{align}
& [p_{a},\dfrac{\partial x^{b}}{\partial q^{\nu}}]p_{b}\dfrac{\partial x^{a}}{\partial q^{\mu}}+p_{a}\dfrac{\partial x^{b}}{\partial q^{\nu}}[\dfrac{\partial x^{a}}{\partial q^{\mu}},p_{b}]\nonumber \\
= & [p_{a},\dfrac{\partial x^{b}}{\partial q^{\nu}}][p_{b},\dfrac{\partial x^{a}}{\partial q^{\mu}}]+[p_{a},\dfrac{\partial x^{b}}{\partial q^{\nu}}][\dfrac{\partial x^{a}}{\partial q^{\mu}},p_{b}]+[p_{a},\dfrac{\partial x^{b}}{\partial q^{\nu}}]\dfrac{\partial x^{a}}{\partial q^{\mu}}p_{b}+p_{b}\dfrac{\partial x^{a}}{\partial q^{\mu}}[p_{a},\dfrac{\partial x^{b}}{\partial q^{\nu}}].
\label{arakalan}
\end{align}
Notice that the first and second terms on the right hand side of (\ref{arakalan}) cancel each other, while the third and fourth terms cancel the seventh and sixth terms of (\ref{pmupnu}) (using (\ref{ozellik})) respectively. Then, what is left is the following: 
\begin{align}
[p_{\mu},p_{\nu}]=\dfrac{\partial x^{b}}{\partial q^{\nu}}[p_{a},p_{b}]\dfrac{\partial x^{a}}{\partial q^{\mu}}+\dfrac{\partial x^{a}}{\partial q^{\mu}}[p_{a},p_{b}]\dfrac{\partial x^{b}}{\partial q^{\nu}}.
\end{align}
One immediately observes that the left hand side is antisymmetric in $\mu $ and $\nu $, while the right hand side is symmetric. The only possibility is then $[p_{\mu},p_{\nu}]=0$.

What is rather interesting about this procedure is that $q^{0}$ and $p_{0}$ commute with every other operator, indicating that uncertainty relations are violated in the normal direction. 

It is now time to derive the Hamiltonian found in \cite{Ogawa}. Remembering that $H=\dfrac{1}{2}p_{a}p^{a}+V(x)$ after Dirac brackets are defined, and $V(x)$ simply becomes $V(q)$ after a general coordinate transformation, one has:
\begin{align}
& H=\dfrac{1}{2}\delta ^{ab}\lbrace \dfrac{\partial q^{\mu}}{\partial x^{a}},p_{\mu} \rbrace \lbrace \dfrac{\partial q^{\nu}}{\partial x^{b}},p_{\nu} \rbrace +V(q),\\
& H-V(q)\equiv K,\\
& K=\dfrac{1}{8}\delta ^{ab}\Bigg(2[\dfrac{\partial q^{\mu}}{\partial x^{a}},p_{\mu}]\dfrac{\partial q^{\nu}}{\partial x^{b}}p_{\nu} + 2p_{\mu}\dfrac{\partial q^{\mu}}{\partial x^{a}}[p_{\nu},\dfrac{\partial q^{\nu}}{\partial x^{b}}] + [\dfrac{\partial q^{\mu}}{\partial x^{a}},p_{\mu}][p_{\nu},\dfrac{\partial q^{\nu}}{\partial x^{b}}] + 4p_{\mu}\dfrac{\partial q^{\mu}}{\partial x^{a}}\dfrac{\partial q^{\nu}}{\partial x^{b}}p_{\nu} \Bigg)\nonumber \\
& = \dfrac{1}{2}p_{i}g^{ij}p_{j} + \dfrac{1}{4}\delta ^{ab}\Bigg([\dfrac{\partial q^{\mu}}{\partial x^{a}},p_{\mu}]\dfrac{\partial q^{\nu}}{\partial x^{b}}p_{\nu} + p_{\mu}\dfrac{\partial q^{\mu}}{\partial x^{a}}[p_{\nu},\dfrac{\partial q^{\nu}}{\partial x^{b}}]\Bigg) + \dfrac{1}{8}\delta ^{ab}[\dfrac{\partial q^{\mu}}{\partial x^{a}},p_{\mu}][p_{\nu},\dfrac{\partial q^{\nu}}{\partial x^{b}}].
\label{K4}
\end{align}
Let us first concentrate on the following expression:
\begin{align}
& [g^{-1/4},p_{i}]=-\dfrac{1}{4}g^{-5/4}\partial _{j}g[q^{j},p_{i}],\\
& \partial _{j}g=gg^{ik}g_{ik,j},\\
& \Gamma ^{i}_{ij}=\dfrac{1}{2}g^{ik}g_{ik,j},\\
\Rightarrow & [g^{-1/4},p_{i}]=-\dfrac{i\hbar}{2}g^{-1/4}\Gamma ^{k}_{ki},
\end{align}
where $g\equiv det(g_{ij})$. Now, consider the fourth term of (\ref{K4}):
\begin{align}
& \dfrac{1}{8}\delta ^{ab}[\dfrac{\partial q^{\mu}}{\partial x_{a}},p_{\mu}][p_{\nu},\dfrac{\partial q^{\nu}}{\partial x_{b}}] \nonumber \\
= &\dfrac{\hbar ^{2}}{8}\dfrac{\partial x^{a}}{\partial q^{\alpha}}\dfrac{\partial x^{b}}{\partial q^{\beta}}g^{\alpha \beta}\dfrac{\partial }{\partial q^{\lambda}}\Big(\dfrac{\partial q^{\mu}}{\partial x^{a}} \Big)(\delta ^{\lambda}~_{\mu}-n^{\lambda}n_{\mu}) \dfrac{\partial }{\partial q^{\sigma}}\Big(\dfrac{\partial q^{\nu}} {\partial x^{b}} \Big)(\delta ^{\sigma}~_{\nu}-n^{\sigma}n_{\nu}),\\
& \dfrac{\partial x^{a}}{\partial q^{\alpha}}\dfrac{\partial q^{\mu}}{\partial x^{a}}=\delta _{\alpha}~^{\mu},\\ 
\Rightarrow & \dfrac{\partial x^{a}}{\partial q^{\alpha}}\dfrac{\partial }{\partial q^{\lambda}}\Big(\dfrac{\partial q^{\mu}}{\partial x^{a}} \Big)=-\dfrac{\partial q^{\mu}}{\partial x^{a}}\dfrac{\partial }{\partial q^{\lambda}} \Big(\dfrac{\partial x^{a}}{\partial q^{\alpha}} \Big)=-\Gamma ^{\mu}_{\lambda \alpha},\\
\Rightarrow & \dfrac{1}{8}\delta ^{ab}[\dfrac{\partial q^{\mu}}{\partial x_{a}},p_{\mu}][p_{\nu},\dfrac{\partial q^{\nu}}{\partial x_{b}}] =\dfrac{\hbar ^{2}}{8}\Bigg(g^{00}\Gamma ^{i}_{i0}\Gamma ^{j}_{j0} +g^{kl}\Gamma ^{i}_{ik}\Gamma ^{j}_{jl}\Bigg).
\end{align}
Now, making use of the above results and the strong equations $q^{0}=0, \; p_{0}=0$, one obtains:
\begin{align}
H=\dfrac{1}{2}g^{-1/4}p_{i}g^{1/2}g^{ij}p_{j}g^{-1/4}+\dfrac{\hbar ^{2}}{8}g^{00}\Gamma ^{i}_{i0}\Gamma ^{j}_{j0}+V(q).
\end{align}

\subsection{Calculations for Ikegami \textit{et. al.}'s Work}
In this treatment, the Lagrangian is (\ref{ikelag}). Momentum variables then appear to be \cite{Ikegami}:
\begin{align}
p_{a}=\dot{x}_{a}-\lambda \partial _{a}f(x),
\end{align}
and the primary constraint is again:
\begin{align}
\phi _{1}\equiv p_{\lambda}\approx 0.
\end{align}
Then, 
\begin{align}
H^{*}=\dfrac{1}{2}\delta ^{ab}(p_{a}+\lambda \partial _{a}f(x))(p_{b}+\lambda \partial _{b}f(x))+V(x)+up_{\lambda}.
\end{align}
There is one secondary constraint in this case:
\begin{align}
\phi _{2}\equiv [p_{\lambda},H^{*}]=\partial _{a}f(x)(p^{a}-\lambda \partial ^{a}f(x))\approx 0.
\end{align}
Notice that we are still at the classical level, so the brackets are ordinary Poisson brackets. $\phi _{1}$ has zero Poisson brackets with both $x^{a}$ and $p_{b}$, and the matrix $\Delta _{kl}$ used in the definition of Dirac brackets is antisymmetric. This means, the term involving $\Delta _{kl}$ in the definition of Dirac bracket is zero for both; the fundamental brackets are then those obtained from ordinary Poisson brackets:
\begin{align}
[x^{a},x^{b}]_{D}=[p^{a},p^{b}]_{D}=0,\; [x^{a},p_{b}]_{D}=\delta ^{a}~_{b}.
\end{align}
Now, solving $\phi _{2}=0$ for $\lambda $, one obtains:
\begin{align}
\lambda =\dfrac{(\partial _{a}f(x))p^{a}}{\partial _{b}f(x)(\partial ^{b}f(x))}.
\end{align}
This gives the following classical Hamiltonian \cite{Ikegami}:
\begin{align}
H=\dfrac{1}{2}p_{a}(\delta ^{ab}-n^{a}n^{b})p_{b}+V(x).
\end{align} 
Let us now concentrate on $T_{n}$.
\begin{align}
T_{n}\equiv & \dfrac{1}{2}\Bigg(An^{a}p_{a}n^{b}p_{b} +Bp_{a}n^{a}p_{b}n^{b}+CP_{a}n^{a}n^{b}p_{b}+ Dn^{a}p_{a}p_{b}n^{b}+ \nonumber \\
& Ep_{a}n^{b}p_{b}n^{a}+Fn^{a}p_{b}n^{b}p_{a}+ Gp_{a}p_{b}n^{a}n^{b}+Hn^{a}n^{b}p_{a}p_{b}\Bigg)\nonumber \\
= & (A+B+C+D+E+F+G+H)p_{a}n^{a}n^{b}p_{b}+ Bp_{a}n^{a}[p_{b},n^{b}]\nonumber \\
+ & A[n^{a},p_{a}]n^{b}p_{b}+F[n^{a},p_{b}]n^{b}p_{a}
+ Ep_{a}n^{b}[p_{b},n^{a}]+ D[n^{a},p_{a}]p_{b}n^{b} \nonumber \\ 
+ & Dp_{a}n^{a}[p_{b},n^{b}] 
+Gp_{a}[p_{b},n^{b}]n^{b}+Gp_{a}n^{a}[p_{b},n^{b}]+Hn^{a}[n^{b},p_{a}]p_{b}\nonumber \\
+ & H[n^{a},p_{a}]n^{b}p_{b}.
\end{align}
Now, let us try to write the geometric potential of da Costa in terms of the commutators of $n^{a}$ and $p_{a}$ (assuming the $q^{3}\rightarrow 0$ limit of the metric given in \cite{Ferrari, daCosta}):
\begin{align}
\partial _{a}n^{a}=\dfrac{\partial q^{\mu}}{\partial x^{a}}\partial _{\mu}n^{a} = \dfrac{\partial q^{i}}{\partial x^{a}}\alpha _{i}~^{j}\dfrac{\partial x^{a}}{\partial q^{j}} + \dfrac{\partial q^{0}}{\partial x^{a}}\partial _{0}n^{a}.
\end{align}
Notice that we used the definition of the Weingarten matrix. Now, one should notice that $\partial _{0}n^{a}$ corresponds to the change in the Cartesian components of $\mathbf{n}$ along the direction of $\mathbf{n}$. Since $\mathbf{n}$ is by definition a unit vector, in the coordinates that are chosen, $\partial _{0}\mathbf{n}=0$. Then:
\begin{align}
\partial _{a}n^{a}=\alpha _{i}~^{i}=Tr\; \alpha .
\end{align}
Using a similar reasoning, one notices that:
\begin{align}
(\partial _{a}n^{b})(\partial _{b}n^{a})= & \dfrac{\partial q^{\mu}}{\partial x^{a}}\partial _{\mu}n^{b}\dfrac{\partial q^{\nu}}{\partial x^{b}}\partial _{\nu}n^{a}=\dfrac{\partial q^{i}}{\partial x^{a}}\partial _{i}n^{b}\dfrac{\partial q^{j}}{\partial x^{b}}\partial _{j}n^{a}\nonumber \\
= & \dfrac{\partial q^{i}}{\partial x^{a}}\alpha _{i}~^{j}\dfrac{\partial x^{b}}{\partial q^{j}} \dfrac{\partial q^{k}}{\partial x^{b}}\alpha _{k}~^{l}\dfrac{\partial x^{a}}{\partial q^{l}}=\alpha _{i}~^{j}\alpha _{j}~^{i} =Tr(\alpha ^{2}).
\end{align}
So, one searches for:
\begin{align}
V_{S}=-\dfrac{\hbar ^{2}}{8}\left(2Tr(\alpha ^{2})-(Tr\; \alpha)^{2}\right)= -\dfrac{1}{8}\left(2[p_{a},n^{b}][n^{a},p_{b}]-[p_{a},n^{a}][n^{b},p_{b}]\right).
\end{align}
Hermiticity requires (\ref{hermite}). Using these conditions, one has:
\begin{align}
2T_{n}=& (2Re(A)+C+D+2Re(E)+2Re(G))p_{a}n^{a}n^{b}p_{b}+ [n^{a},p_{a}]\Big((A+G^{*})n^{b}p_{b}+Dp_{b}n^{b}\Big)\nonumber \\
+ & (A^{*}+D+G)p_{a}n^{a}[p_{b},n^{b}] + (E+G)p_{a}n^{b}[p_{b},n^{a}]\nonumber \\
+ & (E^{*}+G^{*})[[n^{a},p_{b}],n^{b}p_{a}]+ (E^{*}+G^{*})n^{b}p_{a}[n^{a},p_{b}].
\end{align}
First, notice that:
\begin{align}
[[n^{a},p_{b}],n^{b}p_{a}]=n^{b}[[n^{a},p_{b}],p_{a}]
\end{align}
since $n^{b}$ is only a function of coordinates and so is $[n^{a},p_{b}]$. Then:
\begin{align}
[[n^{a},p_{b}],n^{b}p_{a}]=-n^{b}[[p_{a},n^{a}],p_{b}]
\end{align}
since $[p_{a},p_{b}]=0$. Notice here the following:
\begin{align}
n^{b}[[p_{a},n^{a}],p_{b}]=\hbar ^{2}n^{b}\partial _{b}Tr\; \alpha = \hbar ^{2}n^{\mu}\partial _{\mu}Tr\; \alpha = \hbar ^{2}n^{0}\partial _{0}Tr\; \alpha =0,
\end{align}
since $Tr\; \alpha $ is an object which is evaluated at $q^{3}=0$. Now imposing
\begin{align}
A+G^{*}=-D, \; E+G=E^{*}+G^{*}, \; 2Re(A)+C+D+2Re(E)+2Re(G)=1,
\end{align}
one obtains
\begin{align}
T_{n}=\dfrac{1}{2}\Bigg(p_{a}n^{a}n^{b}p_{b}+\hbar ^{2}\Big(D(Tr\; \alpha )^{2}- (E+G)Tr(\alpha ^{2})\Bigg).
\end{align}
This result implies that there are two purely arbitrary constants, $D$ and $E+G$ in this approach, so one has to impose some other conditions on the complex numbers to fix the ordering. Including the number of purely arbitrary coefficients, this is the result obtained by Ikegami \textit{et. al}.

Now, let us consider the ordering hypothesis of Ogawa \textit{et. al}. 
\begin{align}
T_{n}= & \dfrac{1}{2}\lbrace \lbrace n^{a},p_{a}\rbrace \lbrace n^{b},p_{b}\rbrace \rbrace \nonumber \\
= & \dfrac{1}{2}p_{a}p^{a}-\dfrac{1}{4}\Bigg(-p_{a}n^{a}p_{b}n^{b}+p_{a}n^{b}p_{b}n^{a}+ p_{a}p_{b}n^{b}n^{a}\nonumber \\
+ & n^{a}n^{b}p_{a}p_{b}+n^{a}p_{b}n^{b}p_{a}-n^{b}p_{b}n^{a}p_{a} -2n^{b}p_{b}p_{a}n^{a}\Bigg).
\end{align}
Here one expects to find
\begin{align}
D=-\dfrac{1}{4}, \; E+G=-\dfrac{1}{2}.
\end{align}
Here, the corresponding results are:
\begin{align}
D=1, \; E+G=-1,
\end{align}
which mean that this ordering does not give the desired result. 

\begin{acknowledgments}
This work is not the greatest of all in the histroy of physics. However, it required great efforts to be completed. I feel that patience becomes the life of both the student and the supervisor in the course of such a work. I would like to thank to my supervisor Dr. Bayram Tekin for his patience and all sorts of aid he has given during my studies. I would also like to thank the people who shared the load of the work with helpful discussions, Dr. Atalay Karasu, Dr. \"Ozg\"ur Sar\i o\u{g}lu, {\.I}brahim G\"ull\"u, \c{C}a\u{g}r{\i}  \c{S}i\c{s}man, Deniz Olgu Devecio\u{g}lu, Suat Dengiz, Burak {\.I}lhan, Sinan De\u{g}er and Enderalp Yakaboylu. Nader Ghazanfari, K\i van\c{c} Uyan\i k, K\i v\i lc\i m Vural and Ozan Y\i ld\i r\i m deserve special mention, for giving aid concerning different aspects of the process at various moments of the period. I owe special thanks to Deniz  Olgu Devecio\u{g}lu for struggling next to me againts the computers, that is, for technical aid in using Texmaker. My family has a special place in my life, and had a more important place during this period, I would also like to thank them. 
	\\
	
There is one special person whom I haven't mentioned yet. I am greatful to \c{C}a\u{g}layan Arslan, for her patience during the preparation of this work, and for everything we share in our life.       
\end{acknowledgments}

\end{document}